\documentclass[useAMS,usenatbib]{mn2e}

\usepackage[active]{srcltx}
\usepackage{graphicx}
\usepackage{txfonts}
\usepackage{natbib}
\usepackage{multirow}
\usepackage{longtable}

\title[Magnetism and binarity of the Herbig Ae star V380 Ori]{Magnetism and binarity of the Herbig Ae star V380 Ori\thanks{Based on observations obtained at the Canada-France-Hawaii Telescope (CFHT) which is operated by the National Research Council of Canada, the Institut National des Sciences de l'Univers of the Centre National de la Recherche Scientifique of France, and the University of Hawaii}\thanks{Based on observations obtained at the Bernard Lyot Telescope (TBL, Pic du Midi, France) of the Midi-Pyr\'en\'ees Observatory, which is operated by the Institut National des Sciences de l'Univers of the Centre National de la Recherche Scientifique of France.}}
\author[E. Alecian et al.]
{E.~Alecian$^1$$^2$\thanks{E-mail: evelyne.alecian@rmc.ca},
 G.A.~Wade$^1$,
 C.~Catala$^2$,
 S.~Bagnulo$^3$,
 T.~B\"ohm$^4$,
 \newauthor
 J.-C.~Bouret$^5$,
 J.-F.~Donati$^4$,
 C.P.~Folsom$^3$ ,
 J.~Grunhut $^1$ and
 J.D.~Landstreet$^6$, \\
 $^1$Dept. of Physics, Royal Military College of Canada, PO Box 17000, Stn Forces, Kingston K7K 7B4, Canada \\
 $^2$Observatoire de Paris, LESIA, 5, place Jules Janssen, F-92195 Meudon Principal CEDEX, France \\
 $^3$Armagh Observatory, College Hill, Armagh BT61 9DG, Northern Ireland, UK \\
 $^4$Laboratoire d'Astrophysique, Observatoire Midi-Pyr\'en\'ees, 14 avenue Edouard Belin, F-31400 Toulouse, France \\
 $^5$Laboratoire d'Astrophysique de Marseille, Traverse du Siphon, BP8-13376 Marseille Cedex 12, France \\
 $^6$Dept. of Physics \& Astronomy, University of Western Ontario, London N6A 3K7, Canada 
}

\begin{document}

\date{Accepted . Received ; in original form }

\pagerange{\pageref{firstpage}--\pageref{lastpage} \pubyear{2009}}

\maketitle

\label{firstpage}

\begin{abstract}
In this paper we report the results of high-resolution circular spectropolarimetric monitoring of the Herbig Ae star V380 Ori, in which we discovered a magnetic field in 2005. A careful study of the intensity spectrum reveals the presence of a cool spectroscopic companion. By modelling the binary spectrum we infer the effective temperature of both stars: $10500\pm 500$~K for the primary, and $5500\pm500$~K for the secondary, and we argue that the high metallicity ($[M/H] = 0.5$), required to fit the lines may imply that the primary is a chemically peculiar star. We observe that the radial velocity of the secondary's lines varies with time, while that of the the primary does not. By fitting these variations we derive the orbital parameters of the system. We find an orbital period of $104\pm5$~d,  and a mass ratio ($M_{\rm P}/M_{\rm S}$) larger than 2.9. The intensity spectrum is heavily contaminated with strong, broad and variable emission. A simple analysis of these lines reveals that a disk might surround the binary, and that a wind occurs in the environment of the system. Finally, we performed a magnetic analysis using the Least-Squares Deconvolved (LSD) profiles of the Stokes $V$ spectra of both stars, and adopting the oblique rotator model. From rotational modulation of the primary's Stokes $V$ signatures, we infer its rotation period $P=4.31276\pm0.00042$~d, and find that it hosts a centred dipole magnetic field of polar strength $2.12\pm0.15$~kG, with a magnetic obliquity $\beta = 66\pm5^{\circ}$, and a rotation axis inclination  $i=32\pm5^{\circ}$. However, no magnetic field is detected in the secondary, and if it hosts a dipolar magnetic field, its strength must be below about 500~G, to be consistent with our observations. 
\end{abstract}


\section{Introduction}\label{sec:intro}

Today, one of the greatest challenges in stellar physics is to understand the origin of the magnetic fields observed in stars across the H-R diagram. It is now well established that the magnetic fields of the sun and other cool, low-mass stars are generated by a convective dynamo occurring in their external envelope. Their magnetic fields are of complex structure, highly variable on short times-scales, and are strongly correlated with fundamental stellar parameters such as the rotation period, mass, and age, consistent with the dynamo theory \citep{baliunas95,saar96,donati03,donati08}.

Around 5\% of the main-sequence, intermediate-mass A and B stars host organised magnetic fields, with strengths between 300~G and 30~kG, that are stable over tens of years \citep[e.g.][]{auriere07}. These characteristics show no correlation with the mass, age or rotation period of the stars. These qualitative differences suggest a different origin of their magnetic fields. Some authors have proposed the generation of magnetic fields inside the convective core of the intermediate mass stars, with a similar dynamo process as in the low-mass stars, that would diffuse through their radiative envelope to reach the surface. However, this mechanism is not able to reproduce the observed field characteristics, nor can it explain the long-term stability of these magnetic fields \citep{moss01}.

Currently, the favoured scenario for the origin of magnetic fields in A and B type stars is the so-called "fossil field" hypothesis. This model proposes that the magnetic fields observed in the intermediate mass stars are relics, either of the Galactic magnetic field present in the molecular clouds from which the stars formed, or generated by a dynamo during the first stages of stellar formation. Once a star is born, the relic magnetic field would necessarily survive the various stellar evolutionary stages from the pre-main sequence (PMS) to the post-main sequence phases, without significant regeneration. 

%%%%%%%%%%%%%%%

Magnetic fields are observed among molecular clouds with magnetic strengths between 1 and 100~$\mu$G \citep{heiles97} and also among the main sequence (MS) A/B stars. Until recently we were missing information about the magnetic properties of the PMS intermediate mass stars. According to the fossil theory 1-10\% of PMS intermediate mass stars should be magnetic with similar magnetic topology than in the MS A/B stars, and with magnetic strengths consistent with the conservation of the magnetic flux from the PMS to the MS stages.

The Herbig Ae/Be stars have been defined by \citet{herbig60} as emission line stars of spectral type A or B, associated with nebulae, and situated in an obscured region. These characteristics strongly suggest that the Herbig Ae/Be (HAeBe) stars are very young, still completing their PMS phase. We therefore consider that the HAeBe stars are the PMS progenitors of the MS A/B stars. Many authors have predicted the presence of magnetic fields in these stars \citep{catala99,hubrig04,wade07}. Some of them try to detect them, without much success, except in HD 104237 \citep{donati97}.

%%%%%%%%%%%%%%%%

\begin{table*}
\caption{Log of the observations. Columns 1 and 2 give the UT date and the Heliocentric Julian Date of the observations. Column 3 gives the total exposure time. Column 4 gives the peak signal to noise ratio (per ccd pixel, integrated along the direction perpendicular to the dispersion) in the spectra, at the wavelength indicated in column 5. Columns 6 and 7 give the signal to noise ratio in the reconstructed LSD Stokes $V$ profiles of the primary (P) and the secondary (S). Column 8 gives the longitudinal magnetic field of the primary, column 9 gives the rotation phase of the primary as derived in Section 5.2, and columns 10 and 11 give the radial velocities of both components of the system. Column 12 gives the instrument used.}
\label{tab:log}
\centering
\begin{tabular}{lccrcccr@{$\pm$}lcr@{$\pm$}lr@{$\pm$}ll}
\hline
Date   & HJD           & $t_{\rm exp}$ & S/N & $\lambda$ & \multicolumn{2}{c}{S/N (LSD)} & \multicolumn{2}{c}{$B_{\ell}$} & rotation & \multicolumn{2}{c}{${v}_{\rm radP}$} & \multicolumn{2}{c}{${v}_{\rm radS}$} & Instrument\\
UT Time & (2 450 000+) & (s)     &             & (nm) & P & S & \multicolumn{2}{c}{(G)}    & phase & \multicolumn{2}{c}{(km.s$^{-1}$)} & \multicolumn{2}{c}{(km.s$^{-1}$)} & \\
(1) & (2) & (3) & (4) & (5) & (6) & (7) & \multicolumn{2}{c}{(8)} & (9) & \multicolumn{2}{c}{(10)} & \multicolumn{2}{c}{(11)} & (12) \\
\hline
20/02/05  9:32 & 3421.9000 & 4800 & 122 & 809 &  283 &  752 & -165 & 190 & 0.58 & 28.2 &  1.2 & 52.6 &  5.6 & ESPaDOnS \\
26/08/05 14:18 & 3609.09397 & 4800 & 286 & 809 &  948 & 2366 & -501 & 62 & 0.99 & 28.1 &  1.0 & 26.8 &  5.2 & ESPaDOnS \\
09/01/06 11:32 & 3744.98582 & 4000 & 189 & 809 &  564 & 1363 & 410 & 103 & 0.50 & 27.8 &  1.0 & 44.9 &  3.9 & ESPaDOnS \\
12/01/06 12:45 & 3748.03650 & 1600 & 121 & 809 &  362 &  765 & 174 & 168 & 0.21 & 27.6 &  1.2 & 42.5 &  4.8 & ESPaDOnS \\
11/02/06  9:48 & 3777.91178 & 3200 & 193 & 809 &  605 & 1516 & -347 & 98 & 0.13 & 27.7 &  1.2 & 15.5 &  8.6 & ESPaDOnS \\
12/02/06  9:29 & 3778.89836 & 3200 & 193 & 809 &  634 & 1561 & 209 & 92 & 0.36 & 27.6 &  1.1 & 15.9 &  6.9 & ESPaDOnS \\
13/02/06  9:48 & 3779.91116 & 3200 & 149 & 809 &  439 & 1111 & 186 & 129 & 0.60 & 28.1 &  1.1 & 15.5 &  5.9 & ESPaDOnS \\
17/03/07 19:47 & 4177.32449 & 6300 & 163 & 731 &  530 & 1336 & -89 & 96 & 0.74 & 28.2 &  1.1 & 28.6 &  6.7 & Narval \\
06/11/07  2:44 & 4410.61790 & 6300 & 120 & 731 &  381 &  942 & -282 & 161 & 0.84 & 28.0 &  2.0 & 12.8 &  6.6 & Narval \\
06/11/07  4:45 & 4410.70206 & 6800 & 161 & 731 &  529 & 1327 & -143 & 114 & 0.86 & 28.2 &  1.2 & \multicolumn{2}{c}{} & Narval \\
07/11/07  2:12 & 4411.59568 & 6600 & 163 & 731 &  525 & 1323 & -400 & 121 & 0.06 & 27.6 &  1.3 & \multicolumn{2}{c}{} & Narval \\
07/11/07  4:56 & 4411.70953 & 6800 & 138 & 731 &  433 & 1076 & -367 & 140 & 0.09 & 27.6 &  1.5 & 11.3 &  7.4 & Narval \\
08/11/07  2:36 & 4412.61233 & 6800 & 113 & 731 &  349 &  858 & -62 & 172 & 0.30 & 27.4 &  2.1 & 12.5 & 15.0 & Narval \\
08/11/07  4:52 & 4412.70734 & 7500 & 142 & 731 &  441 & 1116 & 84 & 142 & 0.32 & 27.3 &  1.5 & 12.3 &  8.1 & Narval \\
09/11/07  2:59 & 4413.62880 & 7600 & 71 & 731 &  207 &  508 & 723 & 319 & 0.54 & 27.8 &  3.0 &  9.9 & 15.0 & Narval \\
09/11/07  5:06 & 4413.71713 & 6600 & 98 & 731 &  276 &  698 & 698 & 255 & 0.56 & 28.0 &  1.8 & 11.1 & 11.1 & Narval \\
10/11/07  2:52 & 4414.62396 & 6800 & 178 & 731 &  559 & 1408 & 19 & 106 & 0.77 & 28.1 &  1.2 & 11.5 &  6.4 & Narval \\
10/11/07  4:51 & 4414.70630 & 6800 & 156 & 731 &  489 & 1242 & -189 & 128 & 0.79 & 28.2 &  1.3 & 11.7 &  7.2 & Narval \\
11/11/07  3:06 & 4415.63337 & 6920 & 188 & 731 &  607 & 1513 & -336 & 102 & 0.00 & 27.6 &  1.1 & 11.1 &  5.9 & Narval \\
11/11/07  5:06 & 4415.71711 & 6920 & 192 & 731 &  616 & 1555 & -619 & 106 & 0.02 & 27.6 &  1.2 & \multicolumn{2}{c}{} & Narval \\
12/11/07  3:21 & 4416.64386 & 6280 & 184 & 731 &  613 & 1533 & -108 & 107 & 0.24 & 27.1 &  1.2 & \multicolumn{2}{c}{} & Narval \\
12/11/07  5:11 & 4416.72018 & 6280 & 194 & 731 &  605 & 1527 & -16 & 109 & 0.25 & 27.2 &  1.2 &  9.7 &  5.8 & Narval \\
13/11/07  3:15 & 4417.63978 & 6800 & 199 & 731 &  656 & 1645 & 395 & 94 & 0.47 & 27.6 &  1.1 & 11.7 &  5.6 & Narval \\
13/11/07  5:13 & 4417.72210 & 6800 & 192 & 731 &  608 & 1540 & 510 & 111 & 0.49 & 27.7 &  1.2 & 11.0 &  5.2 & Narval \\
\hline
\end{tabular}
\end{table*}

In order to provide new observational constraints on the fossil field hypothesis and explore the magnetic properties of the PMS intermediate mass stars, we performed a survey of $\sim70$ HAeBe stars with the new generation of high-resolution spectropolarimeters: ESPaDOnS, installed at the Canada-France-Hawaii Telescope (CFHT) in Hawaii, and Narval, installed at the Bernard Lyot Telescope (TBL) at Pic du Midi in France (Alecian et al. in prep.). Among this sample, we have discovered 4 new magnetic HAeBe stars, including V380 Ori \citep{wade05,catala07,alecian08a,folsom08}. Our survey has thereby brought the first strong arguments in favour of the fossil theory. In particular, we have concluded that $\sim6$\% of the Herbig Ae/Be stars observed in our survey are magnetic.

To go further, we have characterised the magnetic fields of 3 of these stars (HD 200775, Alecian et al. 2008; HD 190073, Catala et al. 2007; HD 72106, Folsom et al. 2008) in order to compare their magnetic topologies and strengths with those of the MS A/B stars. We have found that the magnetic Herbig Ae/Be stars host large-scale organised magnetic fields, with important dipole components with polar strengths between 300~G and 1.4~kG, in agreement with those of the main sequence stars and with the fossil model.

V380 Ori, of spectral type B9, has been classified as Herbig Ae/Be star by Herbig (1960). This star is associated with many Herbig-Haro objects, with the H{\sc ii} region NGC 1999 in the Orion OB1c association \citep{corcoran95}, and its spectrum contains strong emission lines \citep{finkenzeller84,finkenzeller84b}. \citet{hillenbrand92} classify V380 Ori as a Group I object whose spectral energy distribution (SED) in the infra-red (IR) is similar to those of the low-mass PMS T Tauri stars. The SED is well reproduced with a model of a flat, optically thick accretion disk. These results firmly establish the PMS status of V380 Ori. 

V380 Ori was detected as a rather bright X-ray source by ROSAT \citep{zinnecker94}. Subsequent X-ray grating spectroscopy with Chandra \citep{hamaguchi05} implied a single-temperature plasma with a relatively high temperature, leading these authors to speculate that magnetic activity was occurring in the close environment of the star. However, as pointed out by \citet{stelzer05}, due to the small binary separation of V380 Ori (see Sect. 3.1) Chandra was not able to resolve the system, and the origin of the (apparently single) X-ray source remains obscure.

The magnetic field of V380 Ori was detected during the first run of our Herbig survey, using ESPaDOnS at CFHT \citep{wade05}. This paper reports subsequent observations of V380 Ori acquired between 2005 and 2007, as well as the analysis that we have performed in order to characterise its magnetic field. In Section~\ref{sec:obs} we describe the observations and data reduction. Section~\ref{sec:spI} describes the intensity spectrum and reports the discovery of a spectroscopic companion. In Section~\ref{sec:fp} we determine the fundamental properties of both components, and in Section~\ref{sec:mg} we describe the magnetic field analysis performed on both the primary and secondary stars. Finally a discussion and a conclusion are given in Section~\ref{sec:conc}.

%
%________________________________________________________________

\section{Observations and data reduction}\label{sec:obs}

Our data were obtained using the high resolution spectropolarimeters ESPaDOnS, installed on the 3.6 m Canada-France-Hawaii Telescope, and Narval, installed on the 1.9 m Bernard Lyot Telescope (Donati et al., in preparation) during many observing runs spread over nearly 3 years from Feb. 2005 to Nov. 2007. We obtained 24 circular polarisation (Stokes $V$) spectra covering the entire optical spectrum from 3690 to 10480 \AA,with a resolving power $\lambda/\Delta\lambda=65,000$, and with signal to noise (S/N) ratio between 70 and 290. Table \ref{tab:log} presents the log of the observations.

We used both instruments in polarimetric mode. Each Stokes $V$ observation was divided into 4 sub-exposures of equal time in order to compute the optimal extraction of the polarisation spectra (Donati et al. 1997, Donati et al., in preparation). We recorded only circular polarisation, as the Zeeman signature expected in linear polarisation is about one order of magnitude weaker than circular polarisation. The data were reduced using the "Libre ESpRIT" package especially developed for ESPaDOnS and Narval, and installed at the CFHT and at the TBL (Donati et al. 1997, Donati et al., in preparation). After reduction, we obtained the intensity Stokes $I$ and the circular polarisation Stokes $V$ spectra of the star observed, both normalised to the continuum of V380 Ori. A null spectrum ($N$) is also computed in order to diagnose spurious polarisation signatures, and to provide an additional verification that the signatures in the Stokes $V$ spectrum are indeed real \citep{donati97}.

%
%________________________________________________________________

\section{Properties of the intensity spectrum}\label{sec:spI}

\subsection{V380 Ori: an SB2 system}\label{sec:bin}

\begin{figure}
\centering
\includegraphics[angle=90,width=8.5cm]{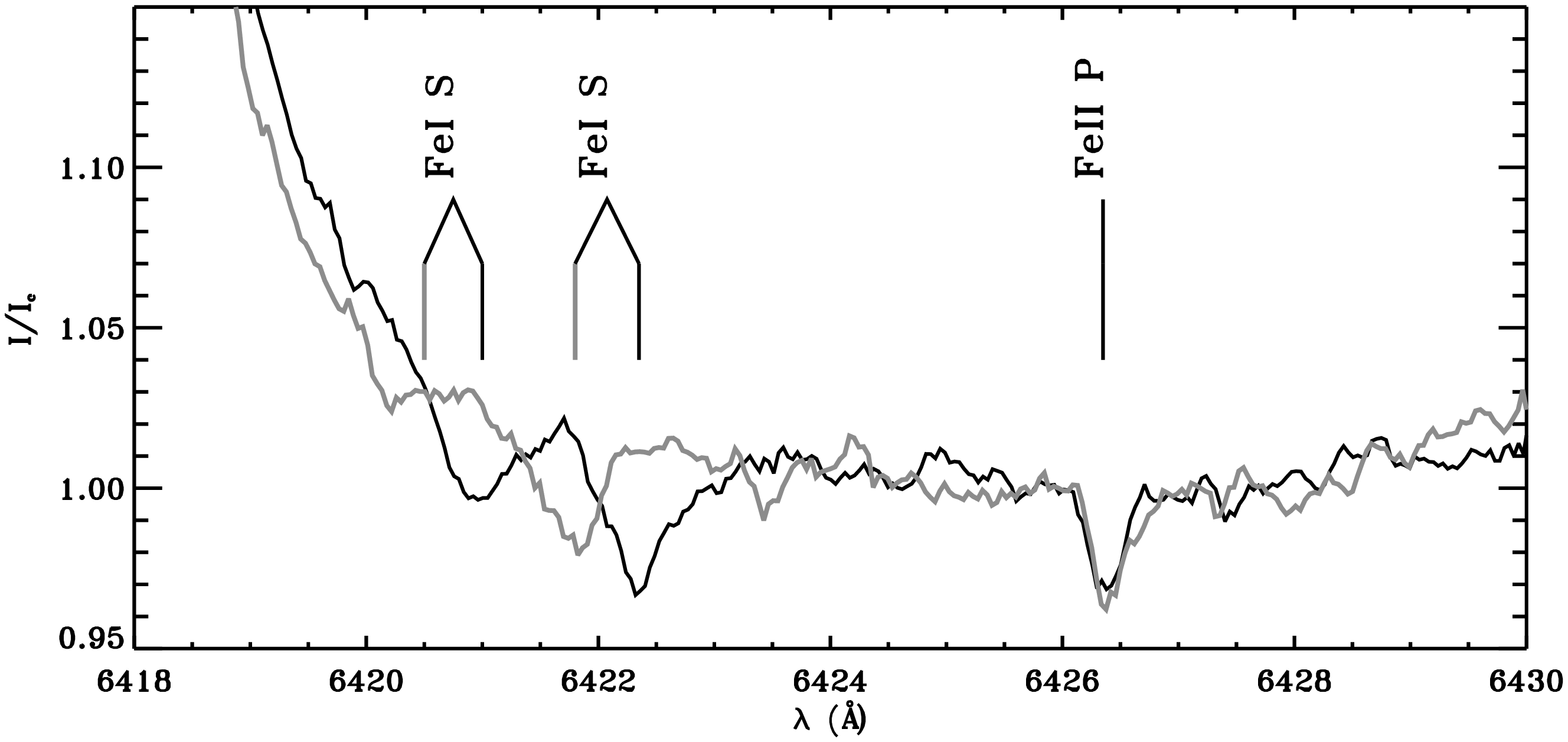}
\includegraphics[angle=90,width=8.5cm]{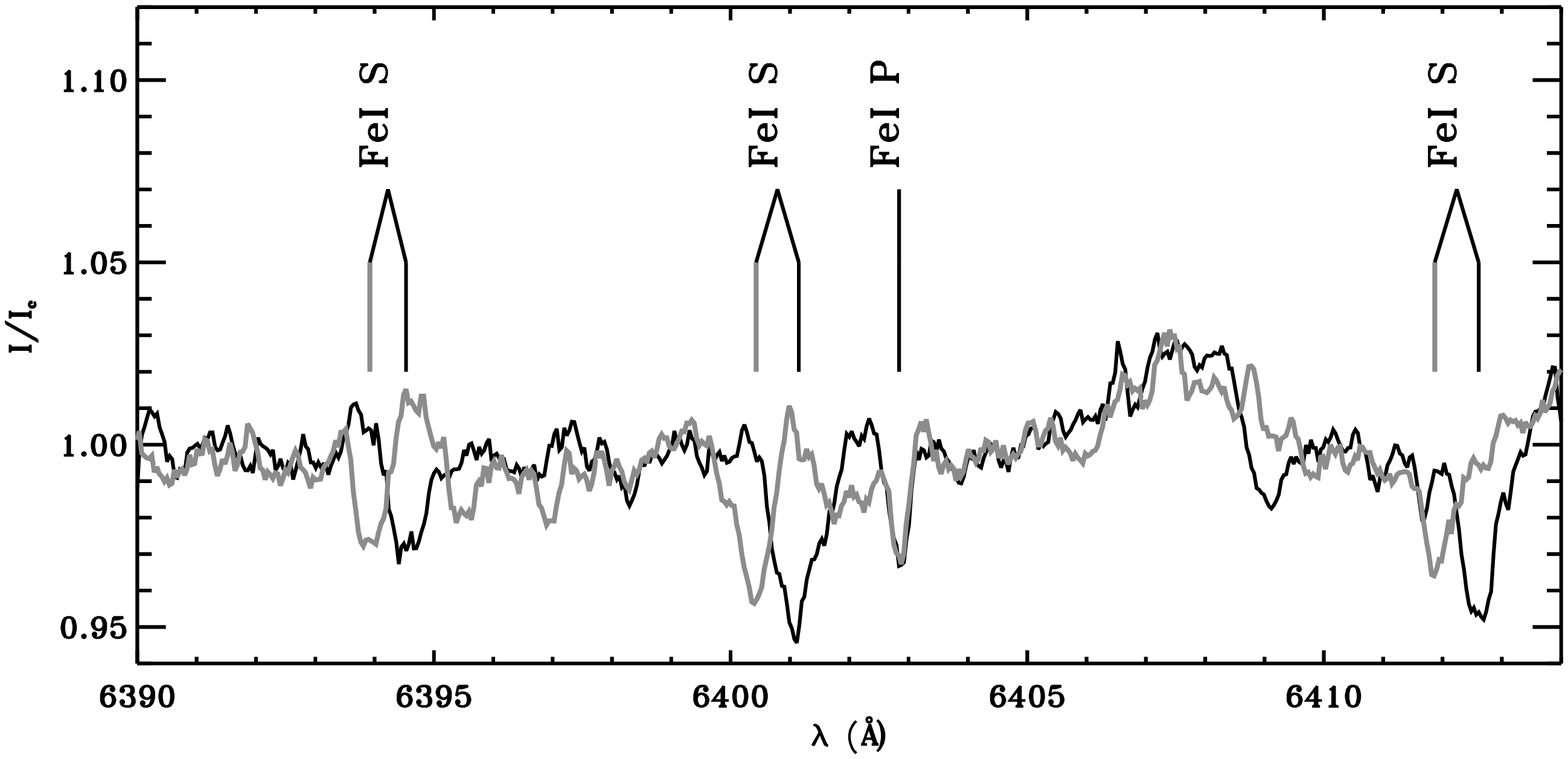}
\caption{Selected regions of the spectrum of V380 Ori observed on January 9th (black line) and February 11th (grey line) 2006 around 6425 \AA\ (up) and around 6400 \AA\ (bottom). The positions and ions of the lines are indicated with a vertical black line for the January spectrum and with a vertical grey line for the February spectrum. The labels P and S beside the ion identification indicate the primary and secondary, respectively. Note the radial velocity variation of the lines of the secondary component, while no shift is observed in the radial velocity of the primary's lines.}
\label{fig:bin}
\end{figure}

Many authors have mentioned the presence of a companion around the variable primary star of V380 Ori. \citet{jonckheere17} was the first to observe a companion of visual magnitude equal to 13~mag at a distance of $3"$ and a position angle of $220^{\circ}$.
However these observations were not confirmed by Herbig (1960), who proposed that Jonckhere was observing bright nebulosities in the vicinity of the central star. 

Later, \citet{leinert94,leinert97} detected an IR companion using speckle-interferometry, a detection that was confirmed by \citet{smith05} and \citet{baines06} using the same technique. The angular separation between the primary and the IR companion is $0.154''$, ie 62~AU at a distance of 400~pc \citep{leinert97}. \citet{bailey98} obtained spectroscopic observations of a sample of PMS stars, including V380 Ori, 2 years after the observations of Leinert (1994,1997), and detected a companion using the spectro-astrometry technique. They measured a position angle of $207^{\circ}$, which is very close to the measurement of Leinert et al. (1997), and they suggested that the fainter component of the system has the greater equivalent width of H$\alpha$. \citet{bouvier01} analysed the system using high-angular resolution spectro-imaging. They detected the companion in December 1993 in only the 3 reddest photometric bands J, H and K. They derived a separation between the primary and the secondary of $0.14''$, and a position angle of $205^{\circ}$. All of these results are generally consistent with the presence of an IR-bright but optically-faint companion star separated from the B9 primary by $\sim 0.15''$.

\citet{corcoran94} detected the Li~{\sc i} 6707~\AA$\,$ line in the spectrum of V380 Ori, confirmed later by \citet{corporon99} and \citet{blondel06}. \citet{corporon99}  attributed the Li line to a companion of spectral type later than F5, but they did not detect any radial velocity variations in the spectral lines of the primary.

In our spectra we confirm the detection of the Li~{\sc i} 6707~\AA$\,$ line, and we detect the presence of other lines belonging to a cool companion (Fig.~\ref{fig:bin}). We observe that the radial velocity of the lines of the cool companion (hereinafter the "secondary"), including the Li~{\sc i} line, vary with a time scale of a month, while the radial velocity of lines of the B9 HAeBe star (the "primary") do not vary detectably with time (see Fig. \ref{fig:bin}). The study of the radial velocities presented in Sect. 4 leads to an estimate of the upper limit of the separation between both stars: $a\sin i < 0.33$ ~AU, which is around 200 times smaller than the separation observed between the IR companion and the primary star by \citet{leinert97}. For these reasons, we conclude that the spectroscopic companion cannot be the IR-bright companion observed using speckle-interferometry, and that V380 Ori is therefore a triple system. In the following the primary (V380 Ori A - the B9 HAeBe star) and the secondary (V380 Ori B - the cooler companion) stars refer to the spectroscopic components, the primary being the hottest and most luminous. The tertiary refers to the IR-bright companion. The large luminosity ratio of the spectroscopic components, their small angular separation ($a\sin i < 0.83$~mas, assuming a distance of 400~pc), and the relatively faint visual magnitude ($V = 10.2$) of the system, explain why the secondary has never been directly detected in the past.

%________________________________________________________________

\subsection{Fitting of the intensity spectrum}\label{sec:fitI}

In order to determine the effective temperatures of both stars and their luminosity ratio, we have fitted the intensity spectrum of V380 Ori with a synthetic spectrum calculated as follows. We used the code BINMAG1 developed by Oleg Kochukhov to calculate the composite spectrum of the binary star. This code takes as input two synthetic spectra of different effective temperatures and gravities, each corresponding to one of the two components. The code convolves the synthetic spectra with instrumental, turbulent and rotational broadening profiles, and combines them according to the radii ratio of the components specified by the user, and the flux ratio at the considered wavelength given by the atmosphere models, to produce the spectrum of the binary star. The individual synthetic spectra were calculated in the local thermodynamic equilibrium (LTE) approximation, using the code SYNTH of \citet{piskunov92}. SYNTH requires, as input, atmosphere models, obtained using the ATLAS 9 code \citep{kurucz93}, and a list of spectral line data obtained from the VALD database\footnotemark\footnotetext{http://ams.astro.univie.ac.at/$\sim$vald/} (Vienna Atomic Line Database).

We compared the observed spectrum of V380 Ori to synthetic binary spectra using effective temperatures ranging from 9500 to 11500 K for the primary and from 4500 to 6500 K for the secondary, and varying the ratio of radii from 1.0 to 1.8. In order to determine the temperature of the primary, we first focussed on regions of the spectrum containing no significant circumstellar emission and lines from the primary not blended with the secondary. Then we varied the temperature of the secondary in order to reproduce all identified lines of the secondary. We find that the observed spectrum is well fit with effective temperatures of $10500\pm 500$ K and $5500\pm500$~K for the primary and the secondary respectively, and with a ratio of radii equal to $1.5\pm0.2$, where the uncertainties are at the $2\sigma$ confidence level.
For both components we used a surface gravity ($g$) given (in cgs units) by $\log g = 4.0$, consistent with the evolutionary stage of these stars. The strong emission contamination does not allow us to measure this characteristic directly. Our spectroscopic determination of the temperature of the primary is consistent with previous determinations of its spectral type, found to be between B9 and A1 \citep{herbig72,finkenzeller84,hillenbrand92}.

\begin{figure}
\centering
\includegraphics[angle=90,width=8.5cm]{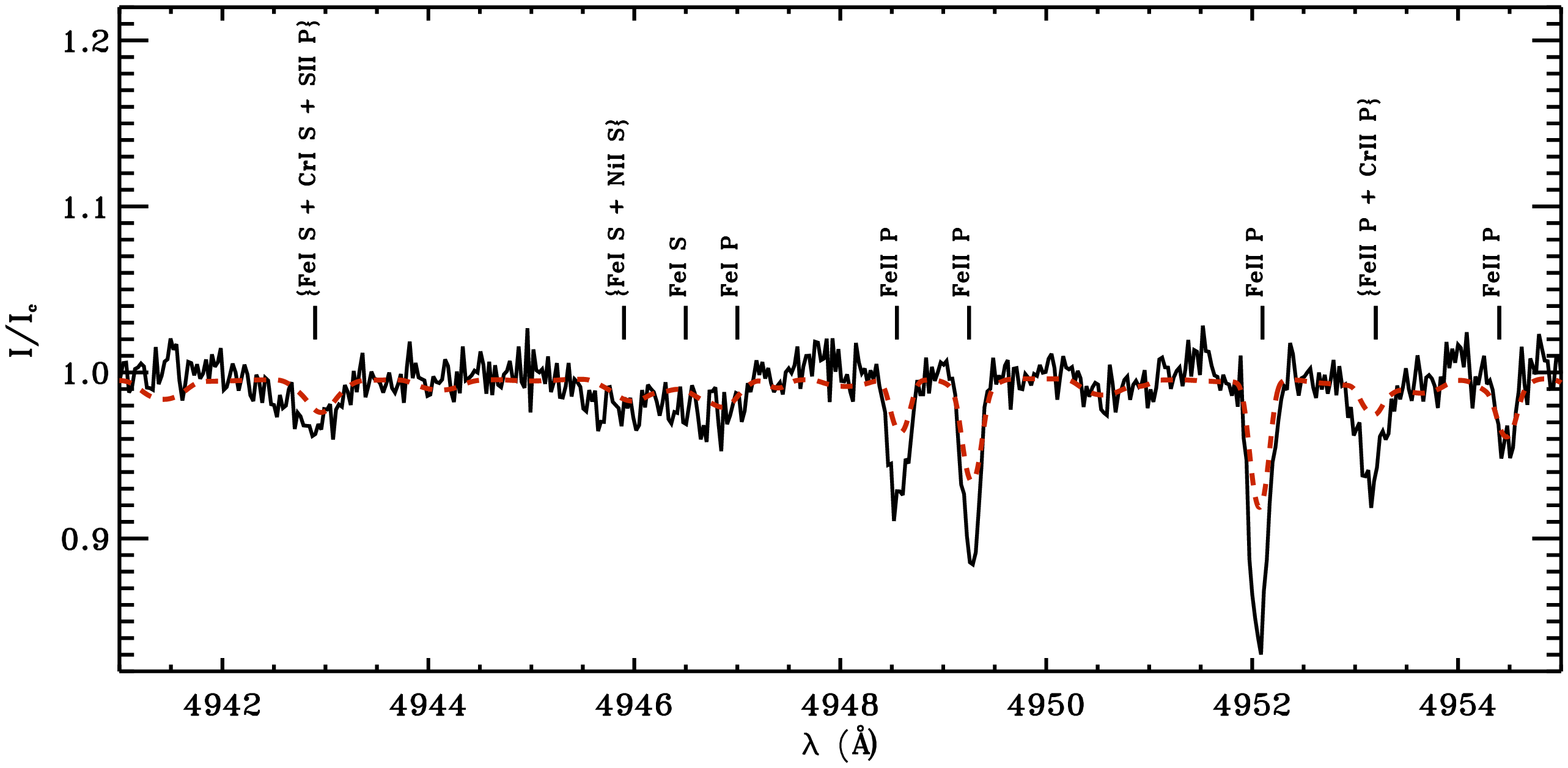}
\hfill
\includegraphics[angle=90,width=8.5cm]{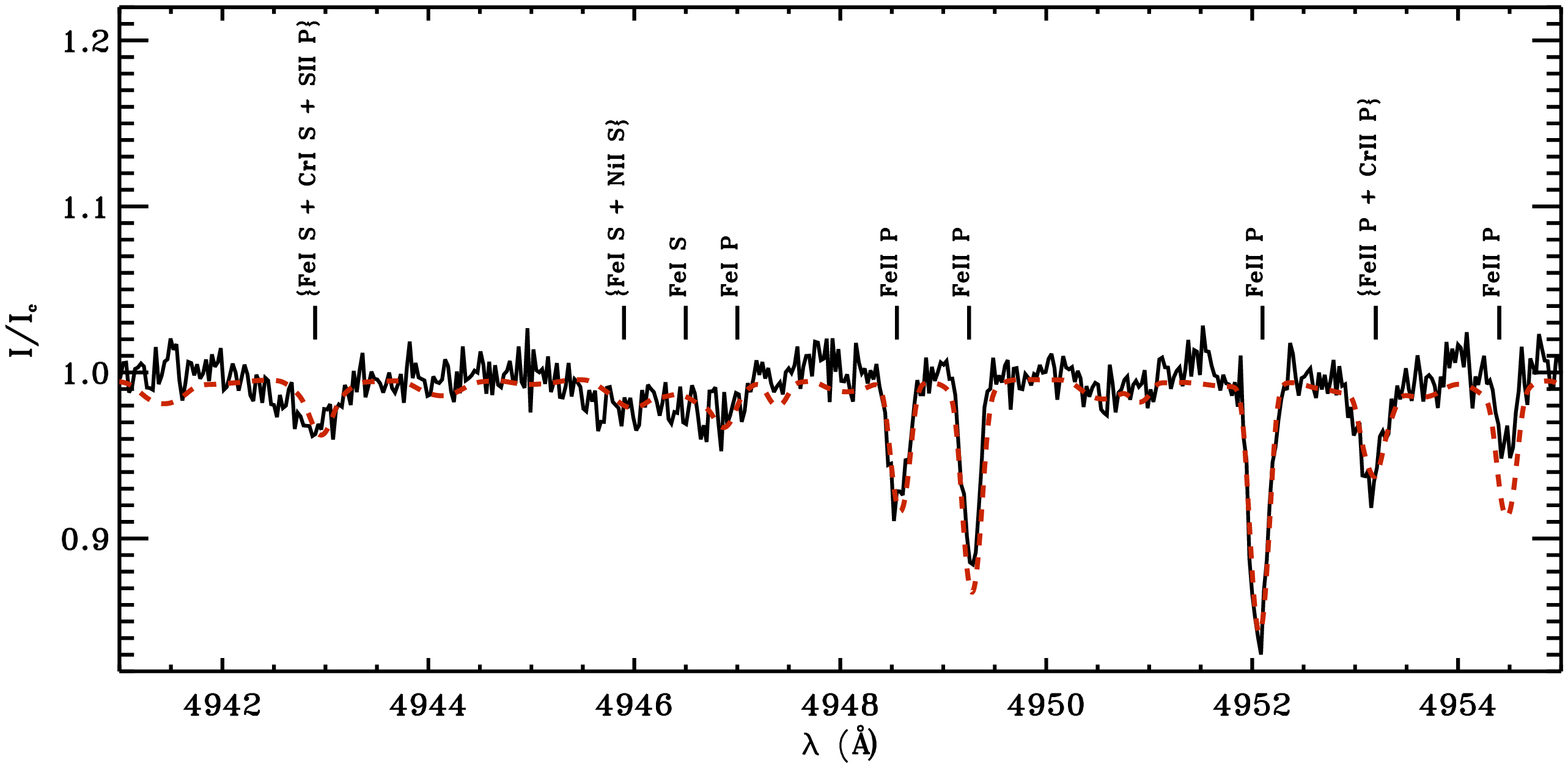}
\caption{Portion of the spectrum observed on August 26th 2005 (black full line) superimposed with the synthetic spectrum of the binary (red dashed line) calculated with a solar metallicity (upper) and a metallicity of $[M/H]=0.5$ (lower). The positions and the ions of the lines are indicated with vertical lines. The labels P and S beside the ion identification indicate lines of the primary and secondary, respectively.}
\label{fig:spmet}
\end{figure}

We also find that the portion of the spectrum not contaminated with emission requires a higher metallicity than the sun in order to adequately fit the lines of the primary. The best-fit metallicity is $[M/H]=0.5$. In Fig. \ref{fig:spmet}, the upper panel shows the observed spectrum superimposed with a synthetic spectrum calculated using BINMAG1 and a solar metallicity, while in the lower panel the synthetic spectrum has been calculated using a metallicity of $[M/H]=0.5$. It is however less evident that the secondary requires a high metallicity in order to fit its lines. While a higher metallicity seems to provide a marginally better fit to the unblended lines of the secondary, the differences between the solar metallicity and the $[M/H]=0.5$ spectra are not sufficiently small as to be insignificant in comparison with the uncertainties of the observed spectrum. 

We can wonder if the high metallicity found in the primary reflects inaccurately determined parameters of the binary system. Using the ratio of radii and the temperatures of both components, we find that the derived luminosity ratio is equal to: $r_{\rm L}=L_{\rm P}/L_{\rm S}=30^{+37}_{-17}$, which implies that the luminosity of the system is dominated by the luminosity of the primary. Therefore increasing $R_{\rm P}/R_{\rm S}$ cannot result in a significant increase in the depth of the synthetic lines of the primary to provide a better fit to the observed lines. Furthermore, we cannot sufficiently strengthen the computed lines of Fe, Ti and Si of the primary by changing the effective temperature or surface gravity within reasonable ranges. We therefore confidently conclude that the metallicity of V380 Ori A is higher than in the sun. 

\begin{figure*}
\centering
\includegraphics[width=16cm]{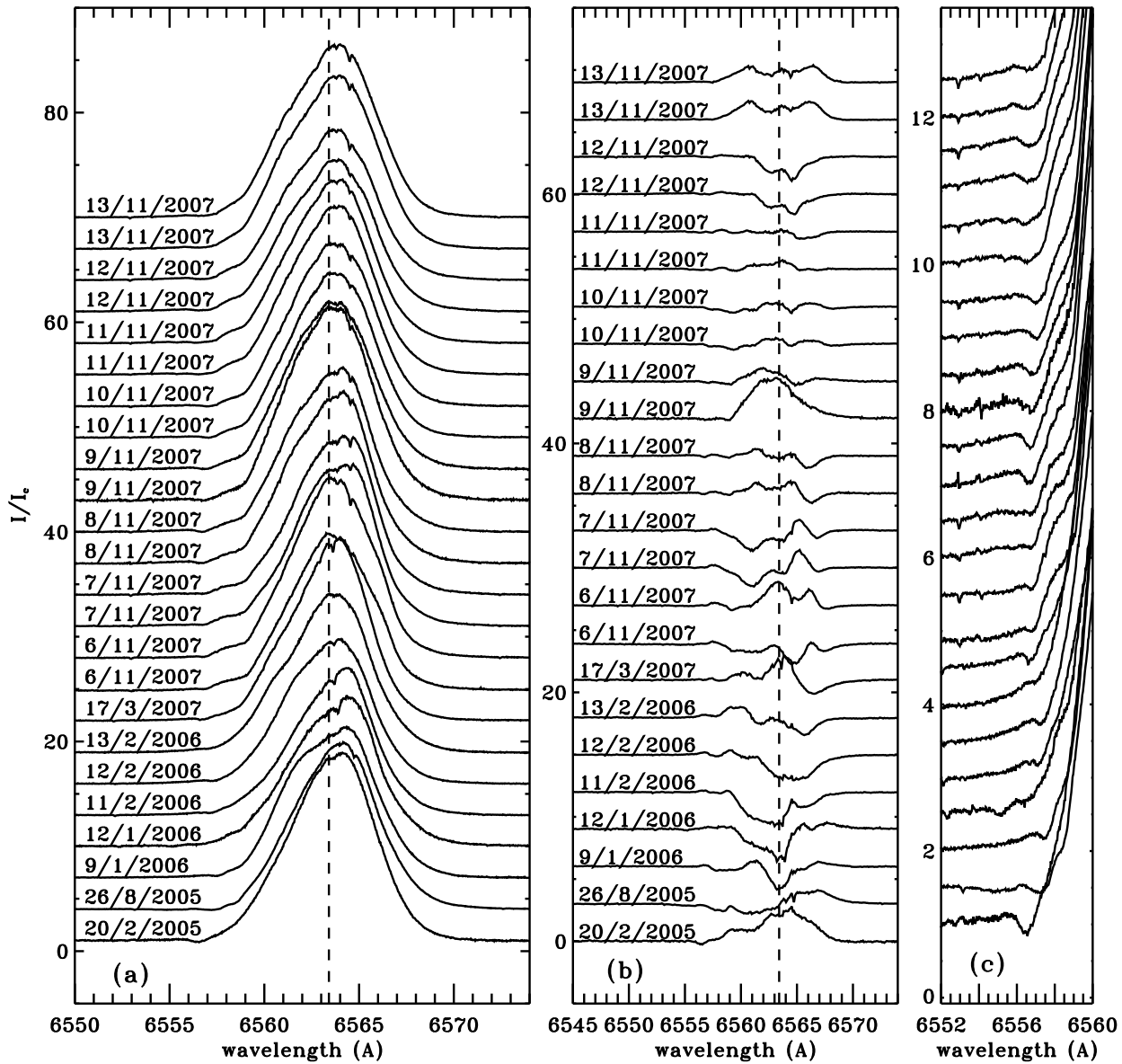}
\caption{{\bf (a)} H$\alpha$ profiles of the 24 observed spectra of V380 Ori. The date is indicated on the left side of the profiles. Note the temporal variations of the amplitude, as well as the appearance and disappearance of substructures in the red and blue wings of the profile. {\bf (b)} Difference between the H$\alpha$ profiles and the mean H$\alpha$ profile. {\bf (c)} Same as (a) but plotted on a different scale to better show the absorption features in the blue wing of the profile.}
\label{fig:ha}
\end{figure*}

It is well known that some magnetic intermediate mass stars on the main sequence show chemical peculiarities characterised by overabundances of elements such as iron, silicon and titanium. These stars are called the magnetic chemically peculiar Ap/Bp stars. First, as the primary of V380 Ori is magnetic, it is reasonable to consider that it might be a chemically peculiar star, or that it may evolve to become one. Secondly, the spectrum of V380~Ori~A is dominated with iron, titanium and silicon lines, and the measurement of the metallicity described above has been mainly performed using these lines. The high metallicity that we measure may therefore reflect only overabundances in iron, titanium and silicon. Thirdly, it seems that the diffusion processes occurring at the surfaces of chemically peculiar stars are sufficiently rapid to observe peculiar stars at the PMS stage of stellar evolution (Vink, Michaud and G. Alecian, private communication). However, at this stage the stars, still surrounded by gas and dust, might experience exchanges of matter with their surroundings, by accretion and winds, which could result in instabilities and in mixing of the stellar atmosphere. The diffusion processes are therefore unlikely to be efficient during the PMS phase. Indeed, we do not observe any chemical peculiarities at the surface of either of the young magnetic HAeBe stars HD 200775 and HD 190073 \citep{alecian08a,catala07}, whereas they are observed in the more evolved star HD~72106 (Folsom et al. 2008) . The strong emission and variability observed in the spectrum of V380~Ori suggests that its environment is highly dynamic, with plenty of gas and dust. However, it is not clear if exchanges of matter occur between the star and its environment. Furthermore the strong magnetic field that we observe at the stellar surface could sufficiently stabilise and isolate its atmosphere, to allow gravitational settling and radiative diffusion to occur. Therefore, we tentatively conclude that the V380 Ori SB2 system is composed of a chemically peculiar primary star and a secondary of solar metallicity. More observations of better S/N ratio, as well as a thorough study of the detailed chemical abundances of both components, are required to confirm this proposal.

%________________________________________________________________

\subsection{The emission lines}\label{sec:emI}

\subsubsection{Description}

The most evident characteristic of the Stokes $I$ spectrum of V380 Ori is a strong contamination by circumstellar emission. All Balmer and Paschen lines visible in the spectrum (from H$\alpha$ to H$\epsilon$, and from P11 to P20), as well as the Ca II 8498 \AA\ and 8662 \AA\ lines are filled with emission. H$\alpha$, H$\beta$, the Paschen lines and the IR Ca II lines exhibit a single-peaked emission profile.  H$\alpha$ shows a faint blueshifted P~Cygni absorption component that may originate from a wind (see Fig. \ref{fig:ha} (a) and (c) for H$\alpha$). The radial velocity of the P~Cygni component changes with time, from -350~km\,s$^{-1}$ to -250~km\,s$^{-1}$, with timescales of the order of few days. Our data reveal no periodicity of these variations. In Fig. \ref{fig:ha} (b) are plotted the differences between the H$\alpha$ profiles and the mean of these profiles, revealing changes in the shape and amplitude of the H$\alpha$ profile on time scales of the order of days and sometimes hours (e.g. between both H$\alpha$ profiles obtained on November 9th 2007 taken two hours apart). Complex structures seem to appear and disappear through the H$\alpha$ profile on timescales of a day. The same kind of structures appear in the H$\beta$ profiles, in the same wing (either blue or red) as the H$\alpha$ profile, but at different radial velocities. 

Most of the metallic photospheric lines in the spectrum of V380 Ori are also superimposed with emission. The emission is correlated with the intrinsic central depth and the excitation potential of the lines: at constant excitation potential, the greater the central depth, the greater the emission, while at constant central depth, the greater the excitation potential, the lower the emission. \citet{catala07} observe in the spectrum of another magnetic HAe star HD 190073, a similar correlation: the emission increases with the central depth of the lines. By re-analysing the spectrum of that star, we find also that at constant central depth: the greater the excitation potential, the lower the emission. These behaviours have also been observed in numerous non-magnetic HAeBe stars (Alecian et al. in prep.). The origin of these correlation is still not understood.

\begin{figure}
\centering
\includegraphics[angle=90,width=8.5cm]{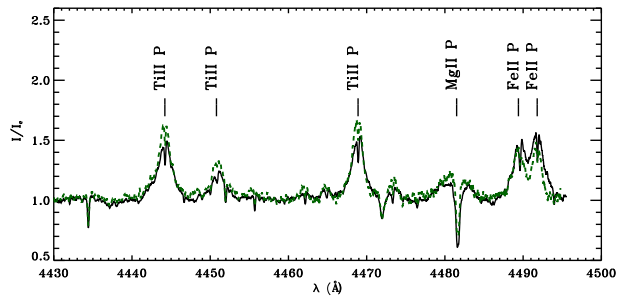}
\caption{Portion of the spectrum observed on August 26th 2005 (black full line) and on February 12th 2006 (green dashed line). The positions and the ions of the lines are indicated with vertical lines. The label P beside the ion identification indicate lines arising from the primary. Note the variation in the emission of the metallic lines of the primary.}
\label{fig:spem}
\end{figure}

\begin{figure}
\centering
\includegraphics[width=8.5cm]{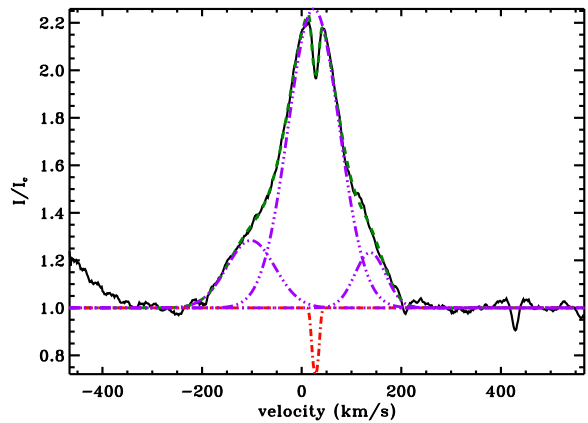}
\caption{Fe~{\sc ii} 4351 \AA\ lines in the spectrum of V380 Ori observed on August 26th 2005 (black full line) superimposed with the fit (green dashed line). The 4 components of the fit, described in Sec. 3.3.1, are also plotted: the 3 Gaussian functions modelling the emission in blue dot-dot-dot-dashed lines, and the photospheric absorption in the red dot-dashed line.}
\label{fig:ironem}
\end{figure}

Most Fe~{\sc ii} and Ti~{\sc ii} lines in the spectrum show strong emission with a presumably photospheric component in absorption, as illustrated in Fig. \ref{fig:spem}. The amplitude and the shape of these emission lines changes on a timescale of days, and sometimes of hours. While the profiles of most of the lines are well reproduced with a simple Gaussian, the strongest emission lines require a slightly more complex model. These emission lines can be described as the superposition of three different Gaussian emissions: a strong main emission component superimposed with two smaller emission, one in each wing of the main component. We have fitted the strongest unblended emission lines using the following model. Each of the 3 emission components is fit using a simple Gaussian with 3 free parameters: the amplitude, the full-width at half-maximum (FWHM), and the centroid. When necessary, we also fitted the photospheric absorption using the convolution of a rotation profile (itself a function of $v\sin i$ and $v_{\rm rad}$) with a Gaussian of instrumental width \citep{gray92}. This convolution is then multiplied by a scale factor in order to fit the depth of the absorption component. An example of the results of this fitting procedure is showed in Fig. \ref{fig:ironem}. While the FWHMs of the 2 wing Gaussians vary from one observation to another, and from one line to another, the strongest emission retains the same value of FWHM, around 65~km\,s$^{-1}$. Temporal variations of the amplitude of the emission, and of the two wing emissions, are observed, but no periodicity is detected. The origin of these emissions is not known. The symmetry of the centroids of the wing emission with respect to radial velocity of the primary suggests that they may arise from a disk surrounding the central star. However the temporal variations of these emissions, i.e. variations of their amplitude, centroid, and FWHM, are difficult to explain in this context.  While the main emission component of the iron and titanium lines is, most of the time, redshifted compared to the radial velocity of the primary star, we observe, in the same spectrum, a few lines for which the main emission component is centred or blueshifted compared to the radial velocity of the star. This behaviour is variable with time, and is not yet understood. Finally, in the few cases in which we include the photospheric component into the fit, the radial velocities and the $v\sin i$ of this component are consistent with the values for the primary's photospheric lines found in Sec. \ref{sec:lsdI}, by fitting the Stokes $I$ LSD profiles.

All the emission lines (except H$\alpha$, H$\beta$, the Paschen lines and the IR Ca II lines) show a photospheric component, as already observed in the spectrum of HD 190073. Furthermore, the characteristic of the main emission component are similar to the emission lines of HD 190073: in both cases the FWHM is around 65 km.s$^{-1}$, constant along the time and through the spectrum. As in Catala et al. (2007) we do not find a suitable scenario to explain these characteristics. These similarities cannot be due to a coincidence, and these emission properties bring therefore strong constraints on the forming region.

Both Ca~{\sc ii} H and K lines show clear P Cygni profiles, as already observed by \citet{herbig60} and \citet{finkenzeller84}. The shape of the emission profile is also composed of a main emission component with wing components similar to those of the Fe~{\sc ii} and Ti~{\sc ii} lines. The radial velocity of the blue absorption part of the P Cygni profiles also varies with time from $-190$ to $-160$~km\,s$^{-1}$ with respect to the rest wavelength of the Ca II lines.

The He~{\sc i} 5876 \AA\ and He~{\sc i} 6678 \AA\ lines show broad emission superimposed with a strong absorption component. Using a Gaussian function we have fitted only the emission component of these lines. We find that the equivalent width of the emission varies from 9 to 27~\AA\ for He~{\sc i}~5876~\AA, and from 10 to 18~\AA\ for He~{\sc i}~6678~\AA, while their FWHMs vary from 130 to 200~km\,s$^{-1}$, and from 130 to 180~km\,s$^{-1}$, respectively. Their centroids vary from -30 to 30~km\,s$^{-1}$, and are, most of the time, blueshifted with respect to the radial velocity of the primary. Interestingly, no emission, and only a small absorption feature is visible at 5876 \AA\ in the the spectrum obtained by \citet{finkenzeller84}. 

The O~{\sc i}~7773~\AA\ triplet is in emission, and superimposed with the photospheric absorptions. Its equivalent width varies with time from 3 to 7~\AA, with temporal variations similar to the He~{\sc i} and Balmer lines, but still without any periodicity.

\subsubsection{Origin of the emission}

The angular separations between the primary and the two companions (0.83 and 154~mas for the secondary and tertiary, respectively) are smaller than the pinhole aperture of the ESPaDOnS instrument (1.6''). Therefore all optical contributions from the 3 components are present in the spectra that we obtained. While no photospheric lines from the tertiary are detected in our spectra, we can wonder if a contribution from this component exists in the emission lines. \citet{leinert97}, using speckle interferometry, measured the flux of the primary and the tertiary separately in 6 IR photometric bands (I', J, H, K, L', and M). They fit the spectral energy distribution using two models: model a) is composed of a system in which the primary dominates the tertiary at all wavelength, whereas model b) assumes an optical primary and an IR companion. In both cases, the primary is dominant at optical wavelengths. It is therefore unlikely that the tertiary contributes to the emission in the optical spectrum of the system.

Excluding the Balmer lines, the major emission in the spectrum is observed only in the lines of the primary - such emission is a common characteristic of HAeBe stars. Our modelling of a few emission lines show that the wing emission components are symmetric with respect to the radial velocity of the primary. Although the main emission component is not perfectly centred on the radial velocity of the primary, this characteristic is commonly observed in other HAeBe stars. Furthermore, the lines of the secondary that are not blended with lines from the primary show emission that is much weaker than that observed in the primary's lines. Finally, the variability observed in the strong emission lines described above has been compared to the radial velocity variations of the secondary, and no correlation has been observed. The emissions observed in the metallic lines of the primary are therefore likely originating from the primary star of the system. However, the small separation ($a\sin i<0.33$~AU) between both spectroscopic component, and the fact that the typical length of the disks surrounding HAeBe stars is larger than few AU \citep{hillenbrand92}, lead to the conclusion that the disk would be circumbinary, instead of circumstellar.

We cannot firmly conclude that this is true for the Balmer lines, and especially H$\alpha$ and H$\beta$. The structure and the variability observed in these lines is complex and our observations cannot rule out that the secondary might contribute to them, and even the tertiary as suggested by the spectro-astrometry study of \citet{bailey98}.

The temporal variations observed among the emission features of the spectrum of V380 Ori are certainly the result of strong and rapid dynamic phenomena in the circumstellar environment of the star, which would be the subject of a very interesting study, but which is not the purpose of this paper. We therefore do not go further in the description and analysis of the emission features observed in the spectrum. Instead, we will look more closely at these characteristics in a future paper that will involve not only high S/N ratio spectroscopic observations, but also other type of observations such as linear polarimetry and interferometry, as well as sophisticated models capable of reproducing the spectroscopic emission of this star, and determining the physical parameters of its environment.

\begin{figure}
\centering
\includegraphics[angle=90,width=8.5cm]{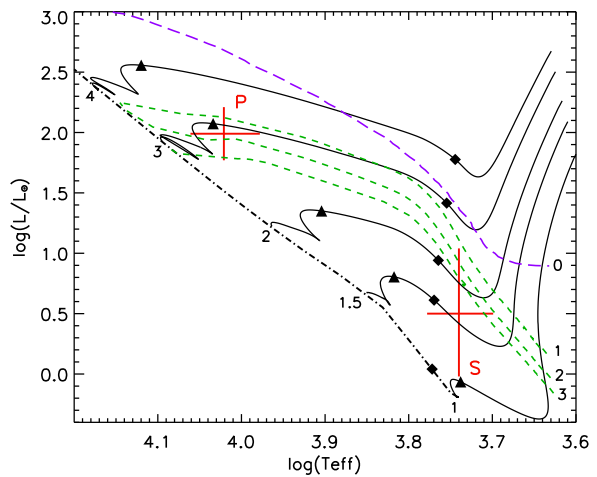}
\caption{HR diagram of V380 Ori. The red crosses are the error bars in temperature and luminosity of the primary (labelled P), and the secondary (labelled S). The PMS evolutionary tracks of 1, 1.5, 2, 3, and 4 $M_{\odot}$ are plotted using full lines. The zero-age main sequence (ZAMS; dot-dashed line), the isochrones (green dashed lines) of 1, 2 and 3 Myr, and the birthline (blue long dashed line), calculated with a mass accretion rate of $10^{-5}$~$M_{\odot}$.yr$^{-1}$ by \citet{palla93}, are also plotted. The filled diamonds indicate the moment where the mass of the convective envelope becomes lower than 1\%  of the stellar mass, and the triangles indicate the appearance of the convective core.}
\label{fig:hr}
\end{figure}

%
%________________________________________________________________

\section{Fundamental parameters of the primary and the secondary}\label{sec:fp}

\subsection{The HR diagram}\label{sec:hr}

V380 Ori belongs to the Ori OB1c association situated at a heliocentric distance of $400\pm40$~pc \citep{herbig88,brown94}. The published photometry shows that it is a variable star \citep[$V$ varying from 10.2 to 10.7,][]{dewinter01}. Its companions and its environment are associated with significant dust and gas, and this could be the cause of the observed variability. In this picture, the variability is due to variable extinction in the circumstellar environment. Therefore, to compute the true luminosity of the V380 Ori system, we have adopted the brightest magnitude in $V$ and $B$, published by \citet{dewinter01}, as the apparent magnitude of the system. This magnitude, in combination with a standard interstellar extinction ($R_{\rm V}=3.1$), as well as the solar absolute magnitude $M_V^\odot=4.75$, allow us to derive the luminosity of the system. We find $\log(L/L_{\odot})=2.00\pm0.20$. From the luminosity ratio determined in Sec. 3.2, we calculate the luminosity of the primary and the secondary: $\log(L_{\rm P}/L_{\odot}) = 1.99 \pm 0.22$ and $\log(L_{\rm S}/L_{\odot}) = 0.50^{+0.54}_{-0.56}$, respectively.

Using these luminosities and the temperatures derived in Sec.~3.2, we compared the positions of both stars in the HR diagram (Fig.~\ref{fig:hr}) with PMS stellar evolutionary tracks of solar metallicity, computed with the CESAM 2K code \citep{morel97}, for different stellar masses. From this comparison we infer the ages, masses and radii of both stars. The error bars in mass and radius are determined by taking ellipses around the error in temperature and luminosity. The age of the system is determined by taking the intersection of the age range of the primary (${\rm age}_{\rm P} = 1.8^{+1.2}_{-1.4}$) and the secondary (${\rm age}_{\rm S} = 7^{+24}_{-6}$), assuming that both stars have the same age. The results are summarised in Table~\ref{tab:fp}.

\begin{table}
\caption{Fundamental parameters of the primary and secondary components of V380 Ori. Columns 2 to 5 give the effective temperature, the luminosity, the mass and the radius of the stars. Column 6 gives the age of the system. The projected rotational velocities ($v\sin i$ - derived in Sect. 4.2) are given in column 7. The macroturbulent velocity ($v_{\rm ma}$ - also derived in Sect. 4.2) of the primary is indicated in column 8. $v_{\rm ma}$ of the secondary cannot be derived using our data. The $2\sigma$ uncertainties are indicated.}
\label{tab:fp}
\centering
\begin{tabular}{@{}l@{\ }r@{$\,\pm\,$}l@{\ }l@{\ }l@{\ }l@{\ }c@{\ }r@{$\,\pm\,$}l@{\ }c@{}}
\hline\hline
Star & \multicolumn{2}{c}{$T_{\rm eff}$}         & $\log(L/L_{\odot})$         & $M/M_{\odot}$              & $R/R_{\odot}$         & age     & \multicolumn{2}{c}{${v}\sin i$} & $v_{\rm ma}$\\
        & \multicolumn{2}{c}{(K)}                         &                                       &                                       &                                 &  (Myr) & \multicolumn{2}{c}{(km.s$^{\rm -1}$)} & (km.s$^{\rm -1}$)\\
\hline
P      &   10500     &500                                  & $1.99\pm0.22$               & $2.87^{+0.52}_{-0.32}$ & $3.0^{+1.1}_{-0.8}$ & \multirow{3}{0.5cm}{$2\pm1$} & 6.7 & 1.1 & $3.5\pm0.8$\\
&  \multicolumn{2}{c}{} & & & & & \multicolumn{2}{c}{}\\
S      &   5500        &500                                   & $ 0.50^{+0.54}_{-0.56}$ & $1.6^{+1.0}_{-0.6}$       & $2.0^{+1.8}_{-0.9}$ &                                & 18.7 & 3.2 & \\
\hline
\end{tabular}
\end{table}

\begin{figure*}
\centering
\includegraphics[angle=90,width=16cm]{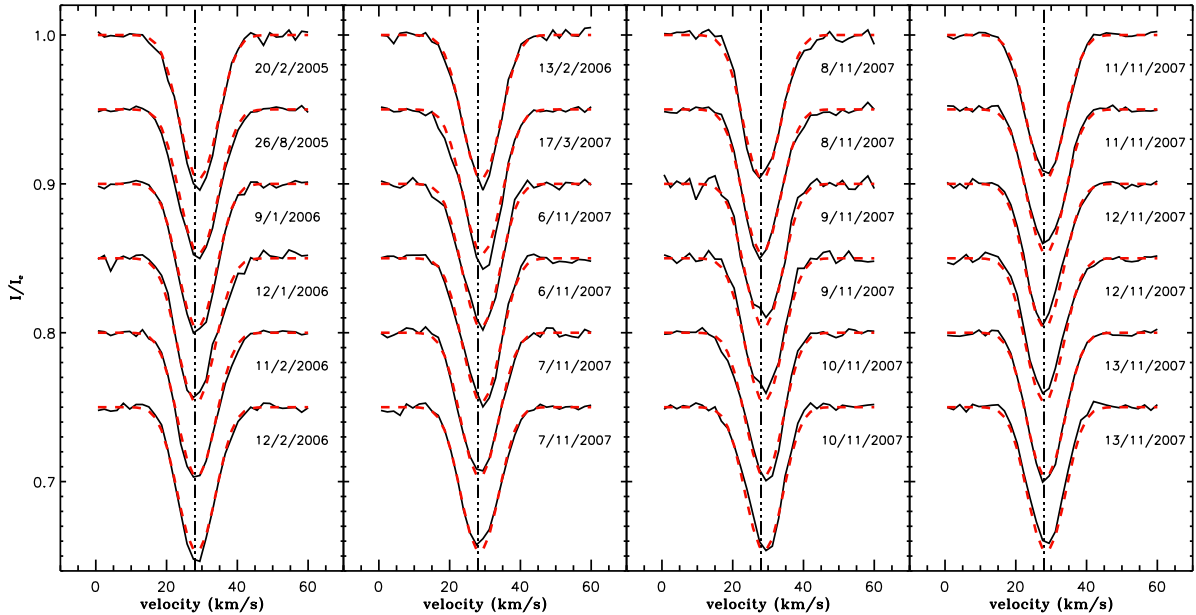}
\caption{LSD Stokes $I$ profiles of the primary star V380 Ori A, calculated from the 24 observed spectra (black full lines), superimposed with the best fit (red dashed line) described in Sec. 4.2. The dates of the observations are indicated beside each profile.}
\label{fig:fitIA}
\end{figure*}

\begin{figure*}
\centering
\includegraphics[angle=90,width=16cm]{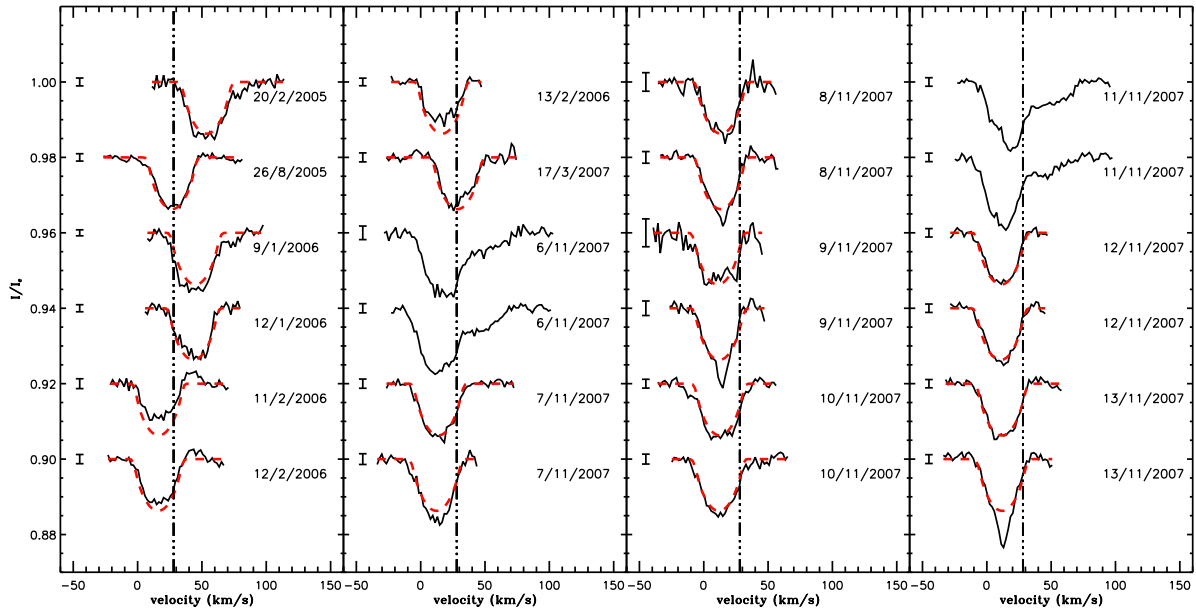}
\caption{LSD Stokes $I$ profiles of the secondary star V380 Ori B, calculated from the 24 observed spectra (black full lines), superimposed with the best fit (red dashed line) described in Sec. 4.2. The dates of the observations are indicated beside each profile. The profiles of November 6th and November 11th 2007 show an absorption component in the red wing, whose the origin is unknown. Therefore these profiles have not been included in the fit.}
\label{fig:fitIB}
\end{figure*}

%________________________________________________________________
\subsection{Calculation and fitting of the LSD Stokes $I$ profiles}\label{sec:lsdI}

In order to increase the signal to noise ratio of our data, to measure accurate rotational and radial velocities of both stars, and to constrain their magnetic field characteristics, we applied the Least Square Deconvolution (LSD) method \citep{donati97} to our spectra using two different line masks (describing the positions, relative intensities and Land\'e factors of all lines predicted to be present in the spectrum), one for the primary component and one for the secondary. Both masks have been calculated using the stellar atmosphere code ATLAS9 of \citet{kurucz93} for the temperatures and gravities of both stars determined in Sect.~3.1. The masks have been cleaned as follows. For each star, we have selected lines that are not blended with the other component at any time, and that are not contaminated by emission, and we have rejected all other lines, including the Balmer lines. The resulting masks contain 85 and 374 lines, respectively for the primary and the secondary component. The number of lines in the mask of the secondary is larger than in the primary mask due to its lower temperature, and also due to the fact that the spectrum of the secondary shows less contamination by emission.

The results of this procedure were two sets of 24 Stokes $I$, $V$ and diagnostic $N$ LSD profiles, for the primary and secondary respectively.The peak signal to noise ratio in the Stokes $V$ profiles, indicated in Table \ref{tab:log}, are higher in the secondary profiles than in the primary's due to a larger number of lines in the secondary mask.

To determine the radial velocity ($v_{\rm rad}$) and the projected rotational velocity ($v\sin i$) of both stars, we performed a simultaneous least-squares fit to the 24 LSD Stokes $I$ profiles, independently for each star. Each profile was fit with the convolution of a rotation function (for which the projected rotational velocity $v\sin i$ is a free parameter in the fit) and a Gaussian whose width is computed from the spectral resolution and a macroturbulent velocity \citep{gray92}.

The free parameters of the fitting procedure are the centroids, depths, $v\sin i$ and macroturbulent velocity $v_{\rm ma}$ of each component. The centroids of the functions can vary from one profile to another, whereas the depths, $v\sin i$ and $v_{\rm ma}$ of each component cannot. This fitting procedure therefore assumes that the depths, $v\sin i$ and $v_{\rm ma}$ of both components do not vary with time, which we confirm (within the error bars) by fitting each profile separately. Including the macroturbulent velocity as a free parameter in the fit of the secondary profiles results in a macroturbulence consistent with zero, but with a very large error bar. As the S/Ns of the secondary profiles are very low, we therefore conclude that this parameter cannot be constrained by fitting these profiles. Therefore, for the fit of the secondary profiles, $v_{\rm ma}$ has not been considered as a free parameter and has been fixed at 0~km.s$^{-1}$.

Fig. \ref{fig:fitIA} and \ref{fig:fitIB} show the resulting fits superimposed on the observed Stokes $I$ profiles of the primary and the secondary components, respectively. The profiles of November 6th and 11th 2007 have not been taken into account in the secondary fit because these profiles contain, in the red wing, a strong absorption component whose the origin is unknown. Including them in the fit results in a poorer fit and in larger error bars. We also observe that the depths of some profiles of the secondary are not well reproduced, which might be due to variable circumstellar emission or absorption contaminating some spectral lines at few observational dates. However, the strongest differences observed between the fit and the observed profiles are within the error bars. Therefore, considering the complexity of the spectrum of the system, the mask used to compute these LSD profiles gives a satisfactory result.

This automatic fitting procedure enables us to measure the $v\sin i$ and radial velocities of both components. We obtain projected rotational velocities of $6.7 \pm 1.1$ and $18.7 \pm 3.2$~km.s$^{-1}$ for the primary and the secondary components respectively, and a macroturbulent velocity for the primary of $3.5 \pm 0.8$~km.s${^-1}$. The radial velocities of both components are included in Table \ref{tab:log}, and are analysed in the next section.

%________________________________________________________________
\subsection{The radial velocity variations}\label{sec:vrad}

The separation between the primary and the tertiary being 200 times larger than the separation between the primary and the secondary, in the following, we will treat the system as a binary system, and we will ignore the impact from the IR companion in the radial velocity analysis.

As mentioned in Sect. 3.3.1, the radial velocities of the spectral lines of the secondary vary with time, while those of the primary do not. The radial velocities of the components, measured using the LSD Stokes $I$ profiles, confirm this observation. We have fitted the temporal variation of the radial velocities of both components assuming a circular orbit. We should note that we also succeeded in fitting these variations using an eccentric orbit. However, the uncertainties are sufficiently large that we have no indication that these orbits are eccentric, and we will only consider the simplest model that can reproduce our measurements. 

\begin{figure}
\centering
\includegraphics[angle=90,width=8.5cm]{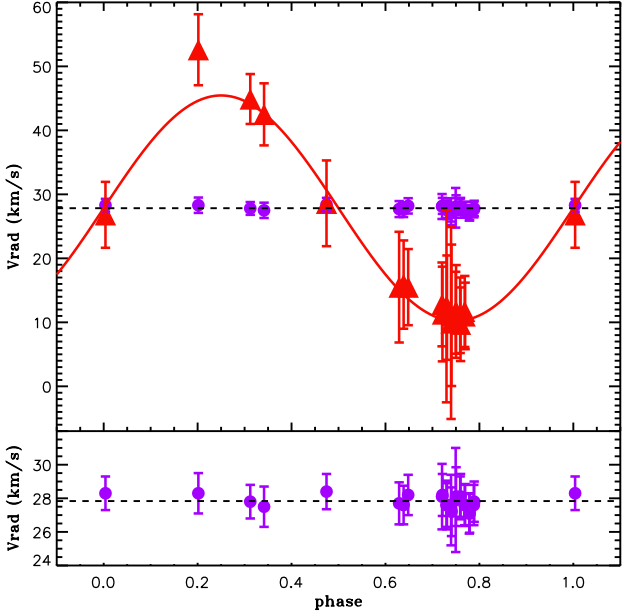}
\caption{Radial velocities of the primary (blue points) and the secondary (red triangle) plotted as a function of the orbital phase (according to the ephemeris given in Table \ref{tab:orb}), and superimposed with the best-fit radial velocity curve of the secondary. The systemic radial velocity is indicated with a black dashed lines. The lower panel shows the radial velocities of the primary on a smaller scale.
}
\label{fig:fitvrad}
\end{figure}

As the radial velocity of the primary does not vary significantly, we chose to include the measurements into the fit as a constant with time equal to the systemic radial velocity, in order to better constrain this parameter. We performed a simultaneous fit of both radial velocity curves using simple sinusoids for each curve. This fitting procedure depends on 4 parameters: the orbital period ($P_{\rm orb}$) of the system, the reference time ($T_{\rm 0}$), the systemic radial velocity ($\gamma$), and the radial velocity amplitude of the secondary ($K_{\rm S}$). We determined an upper limit on the radial velocity amplitude of the primary ($K_{\rm P}$) using the mean of the error bars of the measurement of its radial velocity.

The radial velocities of both components, and the result of the fit of the orbits, are plotted in Fig. \ref{fig:fitvrad}, folded with the orbital period. The value of the fitting parameters are presented in Table \ref{tab:orb}. We derive only a lower limit on the mass ratio $M_{\rm P}/M_{\rm S}$ of 2.9, consistent, within the error bars, with the mass ratio obtained from the HR diagram in Sec. 4.1 ($M_{\rm P}/M_{\rm S} = 1.8^{+1.6}_{-0.8}$). Using the masses derived from the HR diagram, and the inferred value of $M_{\rm P}\sin^3 i_{\rm orb}$ and $M_{\rm S}\sin^3 i_{\rm orb}$, we find that the inclination angle between the orbit axis and the observer's line of site ($i_{\rm orb}$) is lower that $31^{\circ}$, using the primary mass, and lower than $13^{\circ}$, using the secondary mass, which indicates that the orbit might be seen close to pole-on. Due to the relatively low signal to noise ratio of our data, the uncertainties on the orbits and fundamental parameters of the stars are large. More spectroscopic observations of higher signal to noise ratio are required in order to better constrain the orbital parameters and the mass ratio of the system.

\begin{table}
\caption{Orbital parameters of the system. $a$ is the distance between both stars. $a_{\rm P}$ and $a_{\rm S}$ are the radii of the primary and the secondary (circular) orbits, respectively. $i_{\rm orb}$ is the inclination angle between the orbit axis and the line of site of the observer. $K_{\rm P}$ has not been determined within the fitting procedure, but from the error bars of the radial velocities of the primary. The $3\sigma$ error bars are indicated.}
\label{tab:orb}
\centering
\begin{tabular}{ll}
\hline\hline
 & Results of the fit \\
Period (days) & $104 \pm 5$  \\
$T_0$ (HDJ) & $2453505 \pm 18$  \\
$\gamma$ (km/s) & $27.8 \pm 2.1$  \\
$K_{\rm S}$ (km/s) & $18 \pm 14$ \\
$K_{\rm P}$ (km/s) & $<1.4$ \\
 \hline
 & Derived parameters \\
$M_{\rm P}/M_{\rm S}$ & $>2.9$\\
$a\sin  i_{\rm orb}$ (AU) & $< 0.33$  \\
$a_{\rm P}\sin  i_{\rm orb}$ (AU) & $< 0.0139$ \\
$a_{\rm S}\sin  i_{\rm orb}$ (AU) & $0.17\pm0.14$  \\
$M_{\rm P}\sin^3 i_{\rm orb}$ ($M_{\odot}$) & $<0.40$\\
$M_{\rm S}\sin^3 i_{\rm orb}$ ($M_{\odot}$) & $<0.018$  \\
\hline
\end{tabular}
\end{table}

%
%______________________________________________________________

\section{Magnetic field analysis}\label{sec:mg}

\begin{figure}
\centering
\includegraphics[angle=90,width=8.5cm]{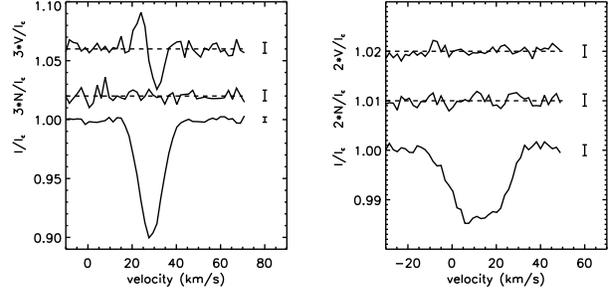}
\caption{LSD Stokes $I$, $V$, and diagnostic $N$ profiles of the primary (left) and the secondary (right), normalised to the continuum of the binary $I_{\rm c}$. These profiles have been obtained from the first November 12th 2007 spectrum. The mean error bars are indicated beside each profile. The $V$ and $N$ profiles have been shifted vertically and have been multiplied by a factor of 3 for the primary and 2 for the secondary, for display purpose.}
\label{fig:ivn}
\end{figure}

Fig. \ref{fig:ivn} shows the LSD Stokes $I$, $V$, and diagnostic $N$ profiles of the primary (left) and the secondary (right) for one night, calculated using the masks described in Sec. \ref{sec:lsdI}. We observe that the null $N$ profile is totally flat in both cases, indicating that the signature observed in the Stokes $V$ profile of the primary is real, and is not a spurious signal created by instrumental artefacts. The magnetic field detection in the primary is furthermore confirmed by the detection of Zeeman signatures in the H$\alpha$ line and in each line of the triplet OI 777 nm \citep{wade05}. Unlike the primary, the Stokes $V$ profile of the secondary doesn't show any Zeeman signature, meaning that no magnetic field has been detected at the surface of the star. It also confirms that the mask of the secondary does not contain lines contaminated with lines from the primary.

In order to characterise the magnetic field of the primary and to put an upper limit on the strength of the magnetic field that the secondary could host, we must perform an analysis of the Stokes profiles of both stars, normalised to their individual continua. The observed spectrum is the combined light coming from both stars, which means that the intensity in the continuum ($I_{\rm c}$) is the sum of the continuum intensity of both stars: $I_{\rm c} = I_{\rm cP}+I_{\rm cS}$. The luminosity of the system being dominated by the luminosity of the primary, the contribution from the secondary's continuum is negligible compared to the error bars in the Stokes $I$ and $V$ profiles of the primary, and won't be taken into account in the analysis of these profiles. However, we cannot neglect the contribution from the primary in the Stokes profiles of the secondary. We have therefore re-normalised these profiles, to the continuum of the secondary only, using the luminosity ratio determined in Sec. \ref{sec:fitI}. As we have been careful to not include lines blended with the companion in the calculation of the LSD profiles, the intensity inside the lines of the secondary is only contaminated with the continuum (and not the lines) of the primary. Therefore the intensity and circular polarisation inside the lines of the secondary are given by:
\begin{eqnarray*}
\frac{I}{I_{\rm c}}=\frac{I_{\rm cP}+I_{\rm S}}{I_{\rm cP}+I_{\rm cS}}=\frac{r_{\rm L}+I_{\rm S}/I_{\rm cS}}{1+r_{\rm L}} \;\;\; {\rm and}  \;\;\; \frac{V}{I_{\rm c}}=\frac{V_{\rm S}}{I_{\rm cP}+I_{\rm cS}}=\frac{V_{\rm S}}{I_{\rm cS}}\times\frac{1}{1+r_{\rm L}} 
\end{eqnarray*}
where $I_{\rm cP}$ and $I_{\rm cS}$ are the continuum intensity of the primary and the secondary respectively, $I_{\rm S}$ is the intensity from the secondary only, $V_{\rm S}$ is the Stokes $V$ profile of the secondary, and $r_{\rm L}$ is the luminosity ratio as defined in Sec. \ref{sec:fitI}. Using these equations, the LSD profiles calculated from the combined spectrum, and the value of $r_{\rm L}$ previously determined, we can extract the Stokes profiles of the secondary normalised to the continuum of the secondary: $I_{\rm S}/I_{\rm cS}$ and $V_{\rm S}/I_{\rm cS}$. In the following we describe the analysis performed on the profiles of both stars.

\subsection{Phased longitudinal field variation of the primary}\label{sec:bl}

In order to characterise the magnetic field of the primary, we need to model the Stokes $I$ and $V$ profiles. The magnetic fields of the Ap/Bp stars and of the other field HAeBe stars are large-scale organised fields, with an important dipole component. We therefore used the oblique rotator model, i.e. a dipole of polar strength $B_{\rm d}$ placed at the centre of the star, with a magnetic axis inclined at an angle $\beta$ with respect to the rotation axis, and the line of sight making an angle $i$ with the rotation axis \citep{landstreet70}. As the star rotates with a period $P$, the magnetic configuration on the observed surface of the star changes with time. Therefore the Stokes $V$ profile, and the magnetic strength projected onto the line of sight and integrated over the observed surface, i.e. the longitudinal magnetic field ($B_{\ell}$), change with time.

\begin{figure}
\centering
\includegraphics[angle=90,width=8cm]{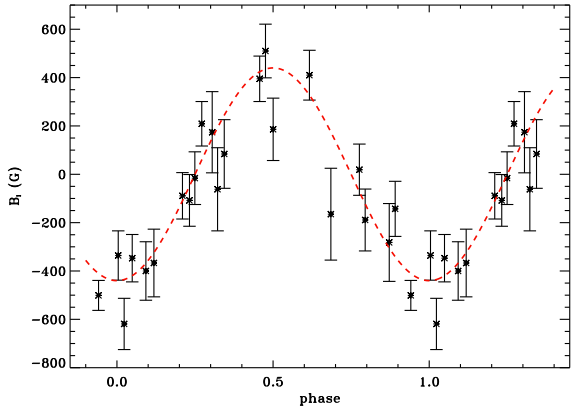}
\caption{Longitudinal magnetic field ($B_{\ell}$) of the primary plotted as a function of rotation phase, according to the ephemeris summarised in Sect. 5.1. The best fit sinusoid is also plotted as a red dashed line. The data points for both November 9th 2007 observations (with relatively large error bars) have been removed from the plot, for display purpose.}
\label{fig:bl}
\end{figure}

As described in Sect. 2, we monitored V380 Ori during many nights spread over about 3 years, in order to fully sample the rotational cycle of the star. We measured the longitudinal magnetic field of the 24 profiles using Eq. (1) of \citet{wade00}:
\begin{equation}
B_{\ell} = -2.14\times10^{12}\frac{\int V(v)vdv}{g\lambda c\int (1-I(v)/I_{\rm c})dv},
\end{equation}
where the wavelength $\lambda$ in \AA, and the Land\'e factor $g$ are the mean values calculated using all the lines of the mask used to compute the LSD profiles, $B_{\ell}$ is in gauss, and $c$ is the speed of light. We calculated the first moment of the Stokes $V$ profile and the equivalent width of the intensity profile, with an integration range from 12 to 45 km\,s$^{-1}$. We found values of $B_\ell$ ranging from $-600$ to $700$~G with uncertainties between 60 and 300~G (see Table \ref{tab:log}). Using a $\chi^2$-minimisation technique, we fit the variation of $B_{\ell}$ with a sinusoidal function of four parameters:
\begin{itemize}
	\item $P$, the rotation period of the star, 
	\item $t_0$, the reference Julian Date of maximum of $B_{\ell}$, 
	\item $B$, the semi-amplitude of the curve, 
	\item $B_{\rm 0}$, the vertical shift of the sinusoidal curve with respect to $B_{\ell}=0$,
\end{itemize}
consistent with the oblique rotator model. The best-fit model, superimposed on the data in Fig. \ref{fig:bl}, gives the following parameters: $P = 4.43\pm0.23$~d, $t_{\rm0} = 2453413.40\pm0.30$~d, $B = 440\pm200$~G, $B_{\rm 0} = 0\pm140$~G, with $\chi^2=1.4$. In contrast, the $\chi^2$ calculated assuming a zero-field model (in which the longitudinal magnetic field is constant and equal to 0) is 11.3. 

The detailed measurements of the longitudinal magnetic field values depend sensitively on the integration range used in the calculation of the first moment of $V$ and the equivalent width. We have measured $B_{\ell}$ for numerous integration ranges, by varying the blue limit from 0 to 18 km,s$^{-1}$, and the red limit from 40 to 60  km.s$^{-1}$. We find that the standard deviation of the values of $B_{\ell}$ measured from a single LSD profile can be as high as 285~G. Furthermore, the measured values for two profiles observed on the same night can show important differences as large as 700 G, while the Stokes $V$ profiles do not show any significant difference. These effects can have important consequences for the determination of the amplitude of variation $B$, as well as the shift of the sinusoid $B_{\rm 0}$. In contrast, however, whatever the integration range used, the $B_{\ell}$ values show similar variations with a periodicity around 4.4~d, implying that the determination of the period is insensitive to the integration range. These characteristics have already been observed during our previous analysis of the magnetic HBe star HD 200775 \citep{alecian08a}. As for HD 200775, we won't try to determine the magnetic geometry of V380 Ori A using the longitudinal magnetic field, and we will determine it by fitting directly the Stokes $V$ variations, using the period derived in this fit as a first guess in the Stokes $V$ modelling. We should note that Fig. \ref{fig:bl} shows that the phase coverage is uniform and complete, which is optimal for modelling the magnetic topology of the star.

%______________________________________________________________

\subsection{Fitting of the Stokes $V$ profiles of the primary}\label{sec:lsdvp}

\begin{figure*}
\centering
\includegraphics[angle=90,width=17cm]{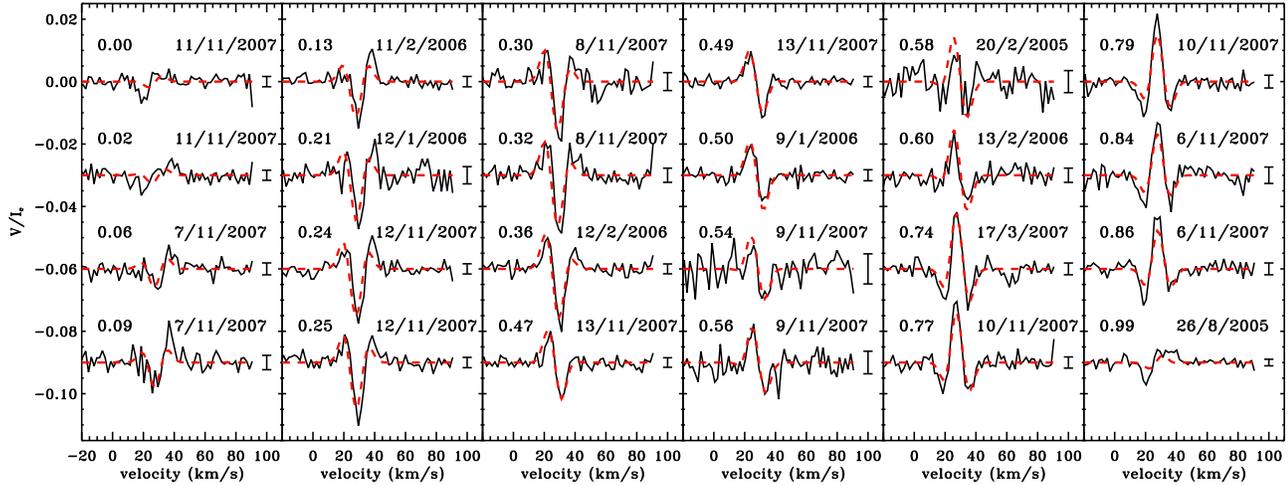}
\caption{LSD $V$ profiles of the primary (full black lines) superimposed by the best oblique rotator model (dashed red line). The numbers close to the profiles are the rotation phase, and the small bars on the right of the profiles are the mean error bars in $V$. The profiles are sorted by increasing rotational phase, and the date of observation is indicated next to each profile.}
\label{fig:fitv}
\end{figure*}

The observed Stokes $V$ profiles are plotted in black full lines in Fig. \ref{fig:fitv}. We see clear variations indicating rotational modulation of the magnetic field at the visible surface of the star. To fit the Stokes $V$ profiles directly, we use the dipole oblique rotator model described above. Using the relations of \citet{landstreet70}, giving the intensity of the magnetic field at each point on the stellar surface, we calculate the longitudinal magnetic field $b_{\ell}(\theta,\varphi)$, at each point ($\theta$,$\varphi$) of the surface of the star, in classical spherical coordinates in the observer's frame.

We assume a gaussian local intensity profile of width $\sigma$ and depth $d$. The width is calculated using the resolving power of the instrument and the macroturbulent velocity determined in Sec.~\ref{sec:lsdI}. The depth is determined by fitting the Stokes $I$ profiles of the primary component, determined in Sec.~\ref{sec:lsdI}. We then calculated the local Stokes $V$ profile at each point on the surface of the star using the weak magnetic field relation of \citet{landi73} : 
\begin{equation}
	V({v},\theta,\varphi)=-C\overline{g}\lambda_0cb_{\ell}(\theta,\varphi)\frac{{\rm d}I(\theta,\varphi)}{{\rm dv}},
\end{equation}
where $C=4.67\times10^{-13}$\AA$^{-1}$G$^{-1}$, $\overline{g}$ and $\lambda_0$ (\AA) are the mean Land{\'e} factor and wavelength of the lines used in the mask (Section 2), $c$ is the speed of light, $I(\theta,\varphi)$ is the local intensity profile, and $v$ is the local radial velocity due to stellar rotation. Then we integrated over the visible stellar surface using the limb darkening law with a parameter equal to 0.5 \citep{claret00}. We obtain the synthetic Stokes $V({v})$ profile, that we normalise to the intensity continuum, to compare to the observed $V$ profiles. This model depends on five parameters : 
\begin{itemize}
	\item $P$ the rotation period
	\item $t_0$, the reference Julian Date of the maximum of the surface magnetic intensity, used with $P$ to compute the rotation phase
	\item $i$, the inclination of the stellar rotation axis to the observers line-of-sight,
	\item $\beta$ the magnetic obliquity angle, 
	\item $B_{\rm d}$ the dipole polar magnetic intensity.
\end{itemize}

We calculated a grid of $V$ profiles for each date of observation (see Table \ref{tab:grid} for details on the grid), varying the five parameters and assuming for the initial value of $P$ the solution obtained from modelling the longitudinal field variation. Then we applied a $\chi^2$ minimisation to find the model which best matches simultaneously our 24 observed profiles. The best model that we found, with $\chi^2=1.6$, corresponds to $P=4.31276\pm0.00042$ d, $i=32\pm5$$^{\circ}$, $\beta=66\pm5$, and $B_{\rm d}=2120 \pm 150$ G, where the error bars correspond to the 3$\sigma$ confidence level. Figure \ref{fig:fitv} shows the synthetic Stokes $V$ profiles superimposed on the observed ones. We see that this model acceptably reproduces most of the observed $V$ profiles.

The period found using this fitting procedure is consistent with that derived from the longitudinal magnetic field variation. However, neither $i$ nor $\beta$ is close to $90^{\circ}$, a result which is inconsistent with the symmetry of the longitudinal magnetic field variation with respect to $B_{\ell}=0$ . However, the amplitude of the Stokes $V$ profiles around phases 0 and 0.5 are clearly different, leading to different absolute values of the longitudinal magnetic fields. From these simple observations we therefore do not expect a symmetric longitudinal magnetic field curve. We are therefore confident with the magnetic configuration found using the fitting of the Stokes $V$ profiles, and we confirm that using $B_{\ell}$ values obtained from low S/N ratio data are unreliable for determining the detailed magnetic configuration of a star.

\begin{table}
\caption{Ranges and minimum bins of parameters explored in the fit of the Stokes $V$ profiles}
\label{tab:grid}
\centering
\begin{minipage}[t]{\linewidth}
\begin{tabular}{llll}
\hline\hline
Parameters & Min & Max & bin \\
\hline
P (days) & 3.0 & 5.0 & 0.00001\\
$T_0$ (HJD) & 2453510 &  2453515 & 0.001\\
$i$ ($^{\circ}$) & 0 & 180 & 0.1\\
$\beta$ ($^{\circ}$) & 0 & 180 & 0.1\\
$B_{\rm d}$ (G) & 0 & 6000 & 10\\
\hline
\end{tabular}
\end{minipage}
\end{table}

\begin{table}
\caption{Magnetic dipole model of V380 Ori A. The $3\sigma$ error bars are indicated.}
\label{tab:mg}
\centering
\begin{minipage}[t]{\linewidth}
\begin{tabular}{ll}
\hline\hline
P (days) & $4.31276\pm0.00042$ \\
$T_0$ (HJD) & $2454412.997\pm0.043$ \\
$i$ ($^{\circ}$) & $32\pm5$ \\
$\beta$ ($^{\circ}$) & $66 \pm 5$ \\
$B_{\rm d}$ (G) & $2120 \pm 150$ \\
\hline
\end{tabular}
\end{minipage}
\end{table}

%______________________________________________________________

\subsection{Stokes $V$ analysis of the secondary}\label{sec:lsdvs}

We used the Stokes $I$ and $V$ profiles of the secondary star to determine the upper limit on the strength of a magnetic field that the star could host. According to the position of the secondary in the HR diagram (Fig. \ref{fig:hr}), its PMS evolutionary status is not well defined. The star might possess a convective envelope under the surface that could give rise to magnetic fields of complex structure, through a convective dynamo as in the T Tauri stars. Therefore we do not claim that this star hosts a large-scale organized magnetic field, however as no magnetic field has been detected we will assume the simplest model, which is the oblique rotator model as described above.

Our method consists of calculating simulated Stokes $V$ observations of the same S/N ratio as our data. For each one of the 24 LSD profile sets, and for a fixed value of the dipole strength $B_{\rm d}$, we have calculated 100 simulated Stokes $V$ profiles as follows. For each of the 100 iterations, we adopted random values of $i$ and $\beta$. We then computed 24 synthetic Stokes $V$ profiles using the oblique rotator model, assigning a random phase to each LSD profile. In this calculation we used a Stokes $I$ profile computed with the $v\sin i$, central depth and macrotubulent velocity found in Sec. \ref{sec:lsdI}, a radial velocity equal to 0 km\,s$^{-1}$, and a limb-darkening coefficient of 0.7 \citep{claret00}. To the computed Stokes $V$ profiles we then added synthetic Gaussian noise consistent with the S/N ratio of the observations. For each of these simulated profiles we calculated the detection probability (Donati et al. 1997). We consider that a magnetic field would have been detected at 3$\sigma$ if the detection probability is greater than 0.9973. We calculated the number of iterations that gives at least one 3$\sigma$ detection among the 24 simulated profiles, and we repeated this procedure for different values of $B_{\rm d}$. We find that if $B_{\rm d}$ is greater than 800 G, the probability to have at least one detection among our data is 99\%, while if $B_{\rm d}$ is greater than 500~G this probability is 90\%. Therefore, if the secondary host a dipole at its centre, its strength is very likely to be below 500~G.

%______________________________________________________________

\section{Discussion and Conclusions}\label{sec:conc}

This paper describes the analysis of high-resolution spectropolarimetric observations of the pre-main sequence Herbig Ae star V380 Ori, obtained with the new generation instruments ESPaDOnS and Narval. The main goal of this study was to characterise the stellar magnetic field discovered in 2005 \citep{wade05}, and to further explore the origin of the magnetic fields of intermediate mass stars. The results of the analysis are summarised and discussed below.

First, a careful investigation of the intensity spectrum of V380 Ori led to the discovery of a cool spectroscopic companion, and confirmed the detection of the Li~{\sc i} 6707 \AA\, line reported by previous authors. We argue that this spectroscopic companion cannot be the infra-red companion discoverd by \citet{leinert94}, which implies that V380 Ori is a triple system with two spectroscopic components and an IR companion. 

The modelling of the binary spectrum allowed the determination of the effective temperatures of both stars, as well as the luminosity ratio of the system, and revealed a high metallicity ($[M/H]=0.5$) of the primary. The same metallicity was also able to reproduce the lines of the secondary, but was not required to obtain an acceptable fit of its spectrum. As a magnetic field has been discovered only in the primary, we therefore proposed that the primary is a magnetic chemically peculiar star, as observed among the magnetic A/B star on the main sequence. No variability is observed in the shape and strength of the lines, implying that any non-uniformities in the surface distribution of Fe and Ti are either too small to be detected, or that these elements are distributed symmetrically about the rotation axis.These results, if confirmed with more accurate observations and a thorough study of the abundances, would give the first Herbig Ae star, whose the environment is highly dense, that is able to develop abundance anomalies at its surface. It would also lead to unique constraints on the physical processes at the origin of these anomalies and the timescales for developing them, as well as constraints on the interaction of the star with its environment in the presence of a strong magnetic field.

The intensity spectrum was found to be strongly contaminated with emission. The emission in the metallic lines is correlated with the central depth and the excitation potential of the lines: at constant excitation potential, the greater the central depth, the greater the emission, while at constant central depth, the greater the excitation potential, the lower the emission. The strongest emission, observed mainly in the Fe~{\sc ii} and Ti~{\sc ii} lines, shows 3 sub-structures, each of which was well reproduced with a Gaussian function. The fit of these lines revealed
\begin{itemize}
\item a main emission component with a constant FWHM of $\sim$65~km.s$^{-1}$ over the spectrum and over time, slightly redshifted compared to the radial velocity of the primary,
\item two wing emissions, one in each wing of the main emission component, whose amplitude and broadening vary with time, that might come from a disk surrounding the central star.
\end{itemize}

We argued that these emissions are very likely produced in a circumbinary disk whose the emission properties are mainly driven by the primary's light, and might not be contaminated with the emission of the secondary. However, the three components of the system might contribute to the strong emission profiles of complex structure and variability observed in the Balmer lines. Our data did not allow us to determine any correlation between the temporal variations of the wing emission components and the rotation period of the primary star. We are therefore unable to assess whether the proposed disk surrounding the primary is produced by magnetically confinement, as proposed by \citet{townsend05} to explain the disk surrounding the hotter main sequence Bp star $\sigma$ Ori E. On the other hand, the lack of veiling (associated with accretion shock hot spots), the lack of correlation between the rotation period and emission variability, and the absence of distortion in the Stokes $I$ profiles (signs of hot or cold spots) suggest that magnetospheric accretion is not occurring around the primary star, as one might expect from its PMS status and the similarity between its emission spectrum and those of the T Tauri stars.

The strong similarity of the characteristics of the emission lines between V380 Ori and HD 190073 is puzzling. If the main emission component is arising in a disk, we would expect to measure different FWHM of their emission profile, due to different inclination angle. The probability to observe two magnetic HAe stars under the same angle being very low, the main emission component might come from another forming region with isotropic properties. As in Catala et al. (2007) we can consider the scenario of a heated region at the base of a wind arising from the photosphere, with turbulent motions. However, the problem encountered with HD 190073 is also encountered with V380 Ori: most of the emission lines are redshifted with respect to the radial velocity of the star, while we would expect them to be blueshifted if they come from a wind. The similarity of the emission spectrum of both stars therefore puts very strong constraints on the physical and geometrical properties of their circumstellar environment.

The P Cygni absorption observed in H$\alpha$ and in the Ca H and K lines revealed the presence of a wind surrounding either the central star, or the central binary. The shape, radial velocity and broadening of all these emissions showed strong variations on time-scales of days, and sometimes hours, suggesting a vigorous dynamic in the environment of the stars. More observations of better S/N ratio coupled with linear polarimetric and interferometric data, and compared to models of circumstellar disks and winds, would be required in order to conclude about the circumstellar geometry and the physical processes involved in the formation and the dynamic of these emissions.

Using the LSD Stokes $I$ profiles of both components we determined their projected rotational velocities and radial velocities. We modelled the variations of the radial velocities of both stars and determined the orbital parameters and the mass ratio of the system. As no variations are observed in the radial velocity of the primary, only a lower limit was determined on the mass ratio, a value that is consistent with the mass ratio determined using the HR diagram.

The LSD Stokes $V$ profiles of the primary show a clear Zeeman signature characteristic of an organized magnetic field at the surface of the star. We found that the oblique rotator model of a centred dipole was able to reproduce acceptably the temporal variations of the Stokes $V$ profiles, implying that the large-scale magnetic field of V380 Ori A is a simple dipole, as observed in the other Herbig Ae/Be stars and in the Ap/Bp stars. The result reinforces arguments in favour of a fossil magnetic origin by showing that there is no exception to the apparent rule that all magnetic intermediate mass stars host large-scale, organised magnetic fields.

The fit of the Stokes $V$ profiles led to an inclination of the rotation axis equal to $32\pm5^{\circ}$. However, assuming rigid rotation and using the derived rotation period of $4.31276\pm0.00042$~d, the radius inferred from the HR diagram position, and the $v\sin i$ measured from the LSD Stokes $I$ profiles, we infer $i=11^{+7}_{-4}$$^{\circ}$, which is inconsistent with the determination from the Stokes $V$ fit. In order to understand this difference, we performed different tests. First, we fixed the inclination angle to $11^{\circ}$, and we performed a new fit on the Stokes $V$ profiles. The result produced synthetic Stokes $V$ profiles inconsistent with the observed ones around phase 0.0. Secondly, the plot of the $\chi^2$ as a function of the period shows several local minima, and therefore we searched for the best magnetic configuration, reproducing the Stokes $V$ profiles around the periods associated with the local minima. In each cases, the synthetic profiles were found to be inconsistent with the observed profiles. However, the magnetic strength inferred when we performed these tests remained very close to our final determination ($B_{\rm d}=2100$~G). We therefore conclude that the magnetic configuration and the rotation period that we find using the fit of the Stokes $V$ profiles is robust and is not the source of the disagreement between both values of the inclination angle. 

This disagreement might comes from the determination of the radius of the primary, inferred from the effective temperature and the luminosity of the star. While we are confident with the determination of the temperature within the error bars, our determination of the luminosity is more debateable. The error bars on the luminosity are determined from the uncertainties on the distance and the luminosity ratio ($L_{\rm P}/L_{\rm S}$). We did not take into account the fact that the star is variable, as well as the uncertainty on the interstellar extinction, which might lead to additional uncertainties. We encountered the same problem during our study of HD 200775: an inconsistency between two different determinations of the inclination \citep{alecian08a}. As for HD 200775, we must conclude that the determination of the luminosity of HAeBe stars, as well as the fundamental parameters using the HR diagram, are highly uncertain, and must be used with caution.

If the luminosity that we determined is overestimated, the inferred mass and radius of the primary might be also overestimated. Using our estimation of the mass and using the PMS stellar evolutionary models, we can determine the radius of the primary when it will reach the ZAMS. Notwithstanding that our estimates of masses and radii of the components are highly uncertain, we can reasonably assume that the ZAMS radius of a 10500 K star is greater than 2~$R_{\odot}$. Using our determination of the radius, and the fact that we might have overestimated it, the upper limit on the actual radius is 4~$R_{\odot}$. We therefore deduce that the ratio of the actual radius and the ZAMS radius is not larger than 2, and can be as low as 1. Using this range of $R/R_{\rm ZAMS}$, we can derive the magnetic strength and the period on the ZAMS.

Assuming magnetic flux conservation during the PMS phase, we predict that, when the primary will reach the main sequence, its magnetic strength will be larger than $\sim2$~kG, which is of the same order of magnitude as the magnetic strengths of the Ap/Bp stars, as expected in the context of the fossil model. Assuming the conservation of angular momentum, we find that the rotation period on the ZAMS can be as low as 1.1~d, similar to the rotation periods of the Ap/Bp stars ($>1$~d), meaning that V380 Ori A might evolve towards the ZAMS with a constant angular momentum. 

Assuming that $i=32^{\circ}$, the true equatorial velocity of the primary is estimated to be 13~km\,s$^{-1}$, showing that V380 Ori A is already a slow rotator at an age around 2~Myr. 
In order to explain the slow rotation observed among the Ap/Bp stars, \citet{stepien00} proposed that magnetic braking occurs during the PMS phase of stellar evolution. As V380 Ori A appears already to have lost significant rotational angular momentum, our results show that if this star experienced a braking mechanism it must have been during the very first stages of the PMS evolution, or before.

Using simulated observations of Stokes $V$ profiles of the secondary, calculated from the oblique rotator model, we determined an upper limit of 500~G (at 90\% confidence) on any dipolar magnetic field that this star may host. If the fossil model is correct, one would expect that a binary system should be composed of two magnetic stars with similar magnetic flux, if both stars have formed from the same molecular cloud (which is a reasonable hypothesis considering the young age of this system). Using the radii determined in Sect. \ref{sec:hr}, we find an upper limit of $145$~kG\,$R_{\odot}^{2}$ on the magnetic flux ($B_{\rm d}\times R^2$) of the secondary, and a lower limit of $100$~kG\,$R_{\odot}^{2}$ for the magnetic flux of the primary, which means that our results are consistent with the fossil field hypothesis. However, the radii of the stars are very poorly constrained. In order to better constrain the fossil model we need to improve the determination of the fundamental parameters of both stars by performing interferometric observations, and measuring separately both luminosities. However, the instrumentation available today is limited to bright magnitudes and won't be able to resolve the stars. In few years, once the efficiency of the interferometers has improved, these observations will be feasible and could be leaveraged to bring new constraints to test the fossil hypothesis.

%
%______________________________________________________________

\section*{Acknowledgments}

We deeply thank Mathieu Vick, Georges Michaud, Georges Alecian and Pierre Kervella for fruitful discussions. We are very grateful to O. Kochukhov, who provided his BINMAG1 code. We really appreciate our exchanges with Jerome Bouvier regarding his results on high-resolution imaging data of V380 Ori. EA is supported by the Marie Curie FP6 program. GAW and JDL acknowledge support from the Natural Science and Engineering Research Council of Canada (NSERC). GAW further acknowledges DND (Canada) Academic Research Programme (ARP). We warmly thank the referee, A. Hatzes, for his judicious comments.

\label{lastpage}

%
%______________________________________________________________

\bibliographystyle{mn2e}
\bibliography{v380ori}

\end{document}